\documentclass[pra,onecolumn,showpacs,nofootinbib,preprint,floatfix]{revtex4}

\usepackage{amsmath,amsfonts,amssymb,bm}
\usepackage{dcolumn}
\usepackage[final]{graphicx}
\usepackage{bm}
\usepackage{comment}
\usepackage{dsfont}
\usepackage{float}
\usepackage{rotating}
\usepackage{leftidx}
\usepackage{bbold}

\newcommand{\be}{\begin{equation}}
\newcommand{\ee}{\end{equation}}
\newcommand{\bea}{\begin{eqnarray}}
\newcommand{\eea}{\end{eqnarray}}
\newcommand{\bit}{\begin{itemize}}
\newcommand{\eit}{\end{itemize}}

\newcommand{\bfl}{\begin{flushright}}
\newcommand{\efl}{\end{flushright}}

\newcommand{\non}{\nonumber \\}
\newcommand{\nonu}{\nonumber}

\newcommand{\ra}{\rangle}
\newcommand{\la}{\langle}

\newcommand{\mus}{\mathbf{u}}
\newcommand{\mvs}{\mathbf{v}}

\newcommand{\mC}{\mathbf{C}}

\newcommand{\mr}{\mathbf{r}}

\newcommand{\Cu}{\mathbf{C_u}}
\newcommand{\Cv}{\mathbf{C_v}}
\newcommand{\mM}{\boldsymbol{\mathcal{M}}}

\newcommand{\mL}{\boldsymbol{\mathcal{L}}}

\newcommand{\mP}{\boldsymbol{\mathcal{P}}}
\newcommand{\mrh}{\boldsymbol{\rho}}

\renewcommand\Re{\operatorname{Re}}
\renewcommand\Im{\operatorname{Im}}

\includecomment{pdffigure}

\begin{document}

\title{Excitation spectra of many-body systems by linear response:
General theory and applications to trapped condensates}

\author{Julian Grond$^1$}
\author{Alexej I. Streltsov$^1$}
\thanks{Corresponding author: Alexej.Streltsov@pci.uni-heidelberg.de}
\author{Axel U. J. Lode$^1$}
\author{Kaspar Sakmann$^2$}
\author{Lorenz S. Cederbaum$^1$}
\author{Ofir E. Alon$^3$}
\affiliation{$^1$ Theoretische Chemie, Physikalisch--Chemisches Institut, 
Universit\"at Heidelberg, 
Im Neuenheimer Feld 229, D-69120 Heidelberg,
Germany}
\affiliation{$^2$ Department of Physics, Stanford University, Stanford, California 94305, USA}
\affiliation{$^3$ Department of Physics, University of Haifa at Oranim, Tivon 36006, Israel}

\date{\today}

\begin{abstract}

We derive a general linear-response many-body theory capable of computing
excitation spectra of trapped interacting  bosonic systems, 
e.g., depleted and fragmented Bose-Einstein condensates (BECs).
To obtain the linear-response equations we linearize the 
multiconfigurational time-dependent Hartree for bosons (MCTDHB) method,
which provides a self-consistent description of many-boson systems 
in terms of orbitals and a state vector (configurations), 
and is in principle numerically-exact. 
The derived linear-response many-body theory, which we term LR-MCTDHB, 
is applicable to 
systems with interaction potentials of general form. 
For the special case of a delta interaction potential 
we show explicitly that the response matrix 
has a very appealing bilinear form, 
composed of separate blocks of submatrices originating 
from contributions of the orbitals,
the state vector (configurations), 
and off-diagonal mixing terms. 
We further give expressions for the response weights and density response. 
We introduce the notion of the type of excitations, 
useful in the study of the physical properties of the equations.
From the numerical implementation of the LR-MCTDHB equations and 
solution of the underlying eigenvalue problem,
we obtain excitations beyond available theories of
excitation spectra, 
such as the Bogoliubov-de Gennes (BdG) equations. 
The derived theory is first applied to study 
BECs in a one-dimensional harmonic potential.
The LR-MCTDHB method contains the BdG excitations and, also, 
predicts a plethora of additional many-body excitations 
which are out of the realm of standard linear response. 
In particular, our theory describes the exact energy of the 
higher harmonic of the first (dipole) excitation
not contained in the BdG theory.
We next study a BEC in a very shallow one-dimensional double-well potential. 
We find with LR-MCTDHB low-lying excitations which are
not accounted for by BdG,
even though the BEC has only little fragmentation and, hence, 
the BdG theory is expected to be valid. 
The convergence of the LR-MCTDHB theory is assessed by 
systematically comparing 
the excitation spectra computed at several different levels of theory.
\end{abstract}


\maketitle 
 
\section{Introduction}

Correlated and fragmented Bose-Einstein condensates (BECs) and their dynamics 
have attracted a lot of interest in the recent years 
\cite{folman:02,grimm:00,schumm:05,albiez:05,levy:07,esteve:08,maussang:10,
buecker:11,will:10,mark:11}. 
Excitation spectra of dilute BECs have been widely studied within 
Bogoliubov theory \cite{bogoliubov:47,dalfovo:99,leggett:01,pethick:08}, 
which assumes the presence of a simple condensate and only 
a small amount of excitations and quantum depletion. 

However, in many situations fragmented BECs 
appear where 
more than one eigenvalue of the reduced one-body density matrix \cite{penrose:56}
is macroscopic \cite{Noz1,Noz2}. 
Quite generally, when increasing interactions, depletion can become macroscopic. 
Xu \emph{et al.} measured \cite{xu:06} a strong quantum depletion, 
which differs from the results of
calculations within the Bogoliubov approach. 
A vast number of examples of fragmented condensates \cite{mueller:06} 
is available. 
A very natural case for fragmentation is provided by 
double-well systems \cite{spekkens:99}, 
which can exhibit 
on one extreme Mott-insulator-like 
fully fragmented states. 
On the other extreme, when the condensates in the wells are fully coherent, 
they can be described by a Gross-Pitaevskii (GP) wave-function. 
In between these two extremes, 
one has non-trivial fragmented 
states \cite{MCHB,sakmann:08}.
In asymmetric double-wells
even a richer spectrum of fragmented states has been found 
 \cite{alon.prl2:05}.
The two-well scenarios
can naturally be extended to few-well systems \cite{streltsov2:04}. 
In optical lattices the superfluid--Mott-insulator phase transition has been demonstrated 
experimentally \cite{greiner:02}. 
Generally, in quantum systems exhibiting translational and rotational symmetry, 
ground-state fragmentation is likely to occur \cite{mueller:06}. 
Experiments demonstrated fragmentation in metastable situations \cite{shin.prl2:04}. 
The failure of GP theory in low dimensions has been 
shown in \cite{kolomeisky:00,petrov:00}. 
Excitation frequencies measured in optical lattices with 
an external harmonic potential could not be understood 
within Bogoliubov theory \cite{tuchman:06}.  

In those situations 
where not even the ground state is of GP type, 
the Bogoliubov-de Gennes (BdG) equations 
cannot give a valid description 
of the excitation spectra \cite{paraoanu:01}. 
Recently, we presented a linear-response theory for completely 
fragmented condensates \cite{grond:12}, 
based on the {\it best-mean-field} (BMF) approach \cite{cederbaum:03,alon.pla:07}, 
which assumes only a single configuration. 
The linear-response theory 
for the best-mean-field (LR-BMF) revealed an energetic 
splitting of excitations for condensates in deep symmetric and asymmetric 
double-well potentials which is absent in the BdG theory. 

Other approaches in the literature are the direct calculation of 
excitations \cite{masiello:05,MCHB} using self-consistent methods
and the application of the Bogoliubov approximation to 
lattice models \cite{paraoanu:01,oosten:01}. 
A multi-mode Bogoliubov approach has been 
developed in \cite{japha:11} 
and extends the range of validity of BdG in double-wells. 
Other authors also apply 
the linear-response approach, e.g., 
for the sine-Gordon model \cite{gritsev:07} 
and the Gutzwiller model \cite{menotti:08,krut:10,snoek:11}.

The LR-BMF theory, as mentioned above, is formulated and applicable to fully fragmented BECs.
Based on a single-configuration ansatz for the ground-state wave-function,
LR-BMF provides orbital excitations only \cite{grond:12}.
In other words, LR-BMF is applicable to BECs in a deep double-well (lattice)
and provides excitations to its higher bands associated with deformation of the ultra-cold cloud. 
In this paper we make a big step forward and 
develop a completely general and systematic linear-response 
theory for general systems,
which can be condensed or fragmented to any degree.
The starting point is represented by the multiconfigurational time-dependent 
Hartree for bosons (MCTDHB) method \cite{streltsov:07,alon:08}, 
which allows to solve, in principle, 
exactly for the dynamics of a many-body system \cite{Benchmarks}.
Clearly, the present theory is far more powerful than LR-BMF \cite{grond:12},
however for the price of utilizing a much larger Hilbert space to 
describe the general ground state.
We term our new theory for excitation spectra LR-MCTDHB. 
 
Until now, a large variety of new phenomena has been found using MCTDHB.
Let us briefly list them. 
A \emph{counterintuitive regime} in dynamical condensate splitting, 
where the final state is not the fragmented ground state, 
but a low-lying excited state which is coherent 
was discovered in \cite{streltsov:07}. 
The decay of self-trapping in a double-well was 
predicted in \cite{sakmann:09}, 
which is in stark contrast to calculations within the 
two-mode GP equation and 
the Bose-Hubbard (BH) model.
The fragmentation \cite{Fragmenton} and loss of coherence \cite{Swift} 
of attractive BECs and 
of repulsive ones
while tunneling through a barrier
to open space \cite{PNAS} have also been described. 
In \cite{MCTDHB_OCT} protocols to optimally split a BEC were addressed
and in \cite{MCTDHB_Iva} the connection between depletion on the many-body
level and wave chaos on the Gross-Pitaevskii level in an expanding BEC
was unraveled.
 
Linear response of this very powerful and numerically-exact method 
allows us to study excitation spectra in a general way, 
which has not been possible until now. 
Apart from presenting the general theory,
we start with a very simple example: a BEC in a harmonic potential. 
Such a system was the first one which 
has been experimentally 
realized \cite{BEC1,BEC2,BEC3} and its excitations 
measured \cite{jin:96,mewes:96,andrews:97}. 
We now have a tool to study excitations in this system for different
interaction strengths and atom numbers in a precise and systematic way. 
We then proceed to a very shallow double-well potential. 
In this regime, 
depletion and correlations start to build up,
but a spatial separation into a left and a right system is not yet present. 
We examine the effect of a very small barrier on the excitation
spectrum of the interacting system. 

The structure of the paper is as follows. 
We first introduce 
the relevant theoretical concepts in Sec.~\ref{sec:the}, 
such as the linear-response theory in Sec.~\ref{sec:lin} and the 
multiconfigurational time-dependent Hartree for bosons method in Sec.~\ref{sec:mctdhb}. 
Sec.~\ref{sec:lin_frag} is devoted to the theory of linear response of MCTDHB. 
In Sec.~\ref{sec:der} the full derivation is presented for general interactions. 
In Sec.~\ref{sec:orb} the derivation is given for the linearization 
with respect to the orbitals, and in Sec.~\ref{sec:ci} 
for the configuration-interaction (CI) coefficients. 
In Sec.~\ref{sec:del} the special case of the commonly-employed delta potential 
is presented,
and in Sec.~\ref{sec:lr_sys_mat} the resulting linear-response equations
are cast into a matrix form and its properties discussed.
Furthermore, Sec.~\ref{sec:den} deals with the derivation 
of the position-space density response, 
and Sec.~\ref{sec:type} with an analysis tool 
named `type of excitation'. 
Illustrative examples in one-dimension 
for the here developed LR-MCTDHB theory 
are contained in Sec.~\ref{sec:app}. 
Firstly, in Sec.~\ref{sec:harm} a BEC in a harmonic potential is studied, 
and secondly in Sec.~\ref{sec:dw} a BEC in a shallow double-well potential. 
The results and physical implications are discussed.
Finally, we conclude and summarize in Sec.~\ref{sec:con}.

\section{Theoretical concepts\label{sec:the}}

\subsection{Linear-response theory\label{sec:lin}}

Excitation spectra of quantum systems are of fundamental and practical interest.
In principle, probing the linear response of an exact model gives access to 
the exact excitation spectrum \cite{FW71,pitaevskii:03}. 
As a consequence, a dynamical probing of excitations via shaking a system 
at different frequencies and observing the response leads to the 
same excitation energies no matter if it is performed in a linear regime, 
meaning a small driving amplitude, or in a non-linear regime. 
The only difference might be the intensity of the response, i.e., 
some excitations are rather dark in the linear regime. 
However, this depends also on the choice of a suitable observable. 
For example, the density response of a BEC might at certain 
frequencies be much stronger in the momentum than in the position space, 
for examples see, e.g., \cite{grond:12}. 

Linear-response theory is thus a very powerful tool since it 
allows to compute the resonance frequencies in a linear driving regime, 
and hence the exact excitation spectra, in a static framework. 
Clearly, the numerical efforts needed
to compute the time evolution 
dynamics at 
many different frequencies is orders of magnitude larger than
that needed for
solving an eigenvalue problem. 

However, a precise and complete spectrum can only be obtained 
within an exact theory. 
For example, the GP equation is only capable of describing the 
spatial dynamics of a condensed system close to the ground state, 
and hence its excitation frequencies are not the same as of the 
full many-body system. 
Nevertheless, this is the standard method for calculating excitation 
spectra of interacting bosons at zero temperature. 
The linear-response equations of GP are usually 
called Bogoliubov-de Gennes equations 
\cite{dalfovo:99,leggett:01,pethick:08,edwards:96,ruprecht:96}. 
Those equations are equivalent to the equations obtained when
the many-body Hamiltonian is treated
in the Bogoliubov approximation \cite{bogoliubov:47,gardiner:97,castin:98}, 
or, up to a small correction, in the random-phase approximation \cite{esry:97}. 
This can be understood by the fact that the hole states in Bogoliubov theory 
lead to a damping of the excitations, 
which is reflected by the fact that the density in linear-response theory is, 
as we will see, proportional to the sum of the response amplitudes $u+v$, 
where $v$ is typically negative and only finite if there is quantum depletion, 
which is a prerequisite for hole states.

We give a short sketch of the derivation of the BdG equations \cite{ruprecht:96}. 
We denote the GP equation as
\be\label{eq:GP1}
i\dot{\phi}=\hat H_{GP}\phi,\quad \hat H_{GP}=\hat h+\lambda|\phi|^2\,,
\ee
where we have the interaction strength $\lambda=\lambda_0(N-1)$
and $N$ is the number of bosons in the BEC.
Further, we denote a small, time-dependent and periodic perturbation 
in the external applied potential as 
$\hat h(\mathbf{r})\rightarrow\hat h(\mathbf{r})+\delta\hat{h}(\mathbf{r},t)$, where:  
\be\label{eq:pert}
\delta\hat{h}(\mathbf{r},t)=f^+(\mathbf{r})e^{-i\omega t}+f^-(\mathbf{r})e^{i\omega t}\,.
\ee
Hereby, the probe frequency $\omega$ and the driving amplitudes $f^{\pm}$ are real. 
Via the ansatz
\be\label{eq:GP_orb}
\sqrt{N}\phi(\mathbf{r},t)=e^{-i\mu t}\left[\sqrt{N}\phi^0(\mathbf{r})+u(\mathbf{r}) 
e^{-i\omega t} +v^*(\mathbf{r}) e^{i\omega t}\right]\,,
\ee 
which is an expansion around the static solution of the GP equation $\phi^0(\mathbf{r})$
($\mu$ is the chemical potential) 
and assuming the response amplitudes $|u\ra$ and $|v\ra$ to be small, 
the following equation is obtained:
\be\label{eq:lr_gp}
\left(\boldsymbol{\mathcal{L}_{BdG}}-\omega\right)\left(\begin{array}{c}|u\ra 
\\ |v\ra\end{array}\right)=\left(\begin{array}{c}-\sqrt{N}f^+|\phi^0\ra 
\\ \sqrt{N}f^-|\phi^{0,*}\ra\end{array}\right)\,.
\ee
We arrive at the linear-response matrix:
\be\label{eq:lrm_gp}
\boldsymbol{\mathcal{L}_{BdG}}=\left(\begin{array}{cc} \hat H_{GP} 
+\lambda|\phi^0|^2-\mu & \lambda(\phi^0)^2\\
-\lambda(\phi^{0,*})^{2} & -(\hat H_{GP} +\lambda|\phi^0|^2-\mu)
 \end{array}\right)\,,
\ee
where $\hat H_{GP}$ is constructed with $\phi^0$.
Here and in the following we use the
abbreviation $(a^b)^\ast \equiv a^{b,\ast}$.
We refer to Eq.~\eqref{eq:lr_gp} (with zero right-hand side) as 
Bogoliubov-de Gennes equations, 
determining the \emph{response frequencies} $\omega_k$ 
and \emph{response amplitudes} $(|u^k\ra,|v^k\ra)^T$. 
From it we obtain response energies and response amplitudes, 
which are independent of the exact shape of the external perturbation.
For the orbitals, 
which depend on the shape 
of the perturbation, 
we insert the energies and amplitudes into Eq.~\eqref{eq:GP_orb} and obtain:
\be\label{eq:response_gp}
\phi(\mathbf{r},t)=e^{-i\mu t}\left\{\phi^0(\mathbf{r})+
\frac{1}{\sqrt{N}}\sum_k\left[ \gamma_k u^k(\mathbf{r}) 
e^{-i\omega t}+\gamma_k^* v^{k,*}(\mathbf{r}) 
e^{i\omega t}\right]/(\omega-\omega_k)\right\}\,.
\ee
The response weights, 
which also depend on the shape of the perturbation, 
are given as
 \be
 \gamma_k=\sqrt{N}\int d\mathbf{r}[u^{k,*}(\mathbf{r})
f^+(\mathbf{r})\phi^0(\mathbf{r})+v^{k,*}(\mathbf{r})
f^-(\mathbf{r})\phi^{0,*}(\mathbf{r})]\,.
 \ee
In the following, a more general linear-response theory will be derived. 
First we discuss the underlying many-body theory
which is general and, in principle, exact.

\subsection{The multiconfigurational time-dependent Hartree for bosons
(MCTDHB) method\label{sec:mctdhb}}

The basic idea of the MCTDHB method is to make an ansatz for the many-body 
state in terms of superpositions of symmetrized states (permanents), 
to account for the symmetry of the bosons. 
The number $M$ of orbitals 
with which the
permanents are constructed 
is chosen at will, e.g., 
(at least) two for a double-well system, etc. 
Then, a time-dependent variational principle is applied, 
and in this fashion working equations are derived for the shape of the orbitals, 
which make up the permanents, 
as well as for the coefficients (or state vector). 
The equations for the orbitals are coupled non-linear equations, 
with nonlinearities depending on the one- and two-body reduced density matrices. 
They are coupled to the equation for the state vector, 
which is governed by the general many-body Hamiltonian in terms of the mode operators, 
with matrix elements depending on the orbitals. 
The set of equations has to be solved simultaneously. 
This constitutes a self-consistent (time-adaptive) 
approach for the 
condensate dynamics \cite{streltsov:07,alon:08}. 
The only approximation involved is the number of modes, 
which can be chosen at will. 
Hence, for a large enough number of modes, 
the method gives exact results \cite{sakmann:09,Benchmarks,3well}. 
In imaginary time-propagation MCTDHB boils down to 
the static, self-consistent multiconfigurational Hartree 
for bosons (MCHB) \cite{MCHB} theory.
The main computational limitation is due to the state vector, 
which grows exponentially with the number of modes. 
Since this approach avoids the explicit construction
of the secular matrix,
the MCTDHB method can be applied to much larger systems 
and/or longer propagation times. 
Even with a smaller amount of time-adaptive modes, 
however, one can capture a larger amount of excitations 
than with fixed orbitals. For example, 
shape oscillations of fragmented condensates, 
analogous to Bogoliubov quasi-particles in a single BEC, 
can easily 
be described within MCTDHB. 

Briefly,
here are the ingredients of the MCTDHB theory \cite{streltsov:07,alon:08}.
The orbital part of the MCTDHB equations can be written in a compact form as
\be\label{eq:basiseq_1}
i\rho_{ij}\frac{\partial}{\partial t}|\phi_j\ra=
\sum_{j=1}^M\left[\hat Z_{ij}-\mu_{ij}(t)\right]|\phi_j\ra\,,
\ee
where
\bea\label{eq:ham_1st}
\hat Z_{ij}=\rho_{ij}\hat h+\sum_{s,l=1}^M\rho_{isjl}\hat W_{sl}\,.
\eea
The general two-body interaction operators read
\be
\hat W_{sl}(\mr)=\int d\mr' \phi_s^{*}(\mr')\hat W(\mr-\mr')\phi_l(\mr')\,.
\ee
The one- and two-body reduced densities \cite{sakmann:08,Lowdin,Yukalov_Book} 
$\rho_{ij}$ and $\rho_{isjl}$ 
are given, respectively, by 
\bea
&&\rho_{ij}=\langle\mathbf{C}|\hat a_i^{\dagger}\hat a_j|\mathbf C\rangle\,,\quad 
\rho_{isjl}=\langle\mathbf{C}|\hat a_i^{\dagger}
\hat a_s^{\dagger}\hat a_j\hat a_l|\mathbf C\rangle\,,
\eea
where $|\mathbf C\rangle$ stands for the coefficients (state vector).

The Lagrange multipliers account for the orthonormality 
of the orbitals
and are given by 
\be
\mu_{ij}(t)=\sum_{k=1}^M\la\phi_j|\hat Z_{il}|\phi_{l}\ra\,.
\ee
The coefficients' equations read
\be\label{eq:TMMC}
i\frac{\partial \mathbf{C}(t)}{\partial t}=\mathcal{H}\mathbf{C}(t)\,.
\ee
The many-body 
Hamiltonian is given as
\be\label{eq:TMMCham}
\mathcal{H}=\sum_{k,q}h_{kq}\hat a_k^{\dagger}\hat a_q+
\frac{1}{2}\sum_{k,s,q,l}\hat a_k^{\dagger}\hat a_s^{\dagger}\hat a_l\hat a_q W_{ksql}\,,
\ee
with the matrix elements
\be
h_{kq}=\int d\mr\phi_k^*(\mr)\hat h\phi_q(\mr)\,,
\ee
and
\be\label{eq:mvs}
W_{ksql}=\int d\mr d\mr' \phi_k^*(\mr)\phi_s^{*}(\mr')
\hat W(\mr-\mr')\phi_q(\mr)\phi_l(\mr')\,.
\ee
In Eqs.~(\ref{eq:TMMC},\ref{eq:TMMCham}) and hereafter 
we follow the notation of Refs.~\cite{3well,Mapping}
in which the second-quantized Hamiltonian operates 
directly on the state vector
$|\mathbf C\rangle$ and re-addresses its components.

\section{Linear-response theory 
for many-body systems: LR-MCTDHB\label{sec:lin_frag}}

\subsection{Derivation\label{sec:der}}

We derive first the linear response of MCTDHB for 
general two-body interactions, 
and specify it later on to delta-interactions. 
The orbitals' linear response is obtained 
by expanding around stationary ones
\be\label{eq:ansatz_0_o}
\phi_i(\mathbf{r},t)\approx\phi_i^0(\mathbf{r})+\delta\phi_i(\mathbf{r},t)\,,
\ee
and the coefficients' linear response is obtained by 
an expansion around a stationary configuration:
\be\label{eq:ansatz_0_c}
\mathbf{C}(t)\approx e^{-i\mathcal{E}^0 t}\left[\mathbf{C^0}+\delta\mathbf{C}(t)\right]\,.
\ee
Here, $\{\phi_i^0(\mr), i=1,\ldots,M\}$ are the orbitals,
$\mathbf{C^0}$ the vector of coefficients, 
and $\mathcal{E}^0$ is the energy of the 
chosen stationary state, 
typically the ground state,
obtained by solving the 
MCHB equations \cite{MCHB}.

\subsubsection{Orbital part\label{sec:orb}}

The orbital part of the MCTDHB equations including 
a perturbation $\delta\hat h(t)$
of the external potential can be written as
\be\label{eq:basiseq_pert}
\sum_{j=1}^M\left[\hat Z_{ij}-i\rho_{ij}
\frac{\partial}{\partial t}-\mu_{ij}(t)\right]|\phi_j\ra=-\sum_{j=1}^M\rho_{ij}
\delta \hat h(t)|\phi_j\ra\,,
\ee
where we explicitly keep the Lagrange multipliers
\be
\mu_{ij}(t)=\sum_{l=1}^M\la\phi_j|\hat Z_{il}+\rho_{il}\delta\hat h(t)|\phi_{l}\ra\,.
\ee
In zeroth order we obtain the orbital part of the MCHB 
equations for the stationary state, which we denote as
\be\label{eq:basiseq_2}
\sum_{j=1}^M\left[\hat Z_{ij}^0-\mu_{ij}^0\right]|\phi_j^0\ra=0\,.
\ee
All quantities herein are obtained from the 
stationary MCHB solution 
$\{\phi_i^0(\mr), i=1,\ldots,M\}$ and $\mathbf{C^0}$.
In first order we arrive at
\bea\label{eq:LR1}
&&\sum_{j=1}^M\left(\hat Z_{ij}^0-i\rho_{ij}^0
\frac{\partial}{\partial t}-\mu_{ij}^0\right)|\delta\phi_j\ra\non
&&+\Biggl\{\sum_{s,j,l=1}^M\rho_{isjl}^0
\left(\int d\mr' \phi_s^{0,*}(\mr')\hat W(\mr-\mr')
\delta\phi_l(\mr')+\int d\mr' \delta\phi_s^{*}(\mr')\hat W(\mr-\mr')
\phi_l^0(\mr')\right)\non
&&+\sum_{j=1}^M\left[\langle\mathbf{C}^0|
\hat a_i^{\dagger}\hat a_j|\delta\mathbf C\rangle+\langle\delta\mathbf C|
\hat a_i^{\dagger}\hat a_j|\mathbf{C}^0\rangle\right]\hat h \non
&&+\sum_{s,j,l=1}^M\left[\langle\mathbf{C}^0|
\hat a_i^{\dagger}\hat a_s^{\dagger}\hat a_j\hat a_l|\delta\mathbf C\rangle
+\langle\delta\mathbf{C}|\hat a_i^{\dagger}\hat a_s^{\dagger}
\hat a_j\hat a_l|\mathbf C^0\rangle \right]\hat W_{sl}^0\Biggr\}|\phi_j^0\ra\non
&&-\sum_{j=1}^{M}\left[\mu_{ij}^0|\delta\phi_j\ra+
\delta \mu_{ij}(t)|\phi_j^0\ra\right]=
-\sum_{j=1}^M\rho_{ij}^0\delta \hat h(t)|\phi_j^0\ra\,.
\eea
The perturbed Lagrange multipliers are given as
\bea\label{eq_app:muij}
\delta\mu_{ij}(t)&=&\sum_{l=1}^M\delta\left[\la\phi_j|
\left(\hat Z_{il}-i\rho_{il}\frac{\partial}{\partial t}\right)|\phi_l\ra\right]+
\sum_{l=1}^M\rho_{il}^0\la\phi_j^0|\delta\hat h(t)|\phi_{l}^0\ra\non
&=&\sum_{l=1}^M\la\delta\phi_j|\hat Z_{il}^0|\phi_l^0\ra+
\la\phi_j^0|\sum_{l=1}^M\delta\left(\hat Z_{il}|\phi_l\ra\right)+
\sum_{l=1}^M\rho_{il}^0\la\phi_j^0|\delta\hat h(t)|\phi_{l}^0\ra\non
&=&\sum_{l=1}^M\mu_{il}^0\la\delta\phi_j|\phi_l^0\ra+
\la\phi_j^0|\sum_{l=1}^M\delta\left(\hat Z_{il}|\phi_l\ra\right)+
\sum_{l=1}^M\rho_{il}^0\la\phi_j^0|\delta\hat h(t)|\phi_{l}^0\ra\,,
\eea
where we used partial integration in the last step. 
We arrive at projected response equations
\bea\label{eq:lr_orb}
&&\sum_{j=1}^M\left[\hat P\left(\hat Z_{ij}^0-\mu_{ij}^0\right)-
i\rho_{ij}^0\frac{\partial}{\partial t}\right]|\delta\phi_j\ra+
\hat P\Biggl\{\sum_{s,j,l=1}^M\rho_{isjl}^0\delta\hat W_{sl}+
\sum_{j=1}^M\delta \rho_{ij}\hat h
+\sum_{s,j,l=1}^M\delta\rho_{isjl}\hat W_{sl}\Biggr\}|\phi_j^0\ra \non
&&=-\hat P\sum_{j=1}^M\rho_{ij}^0\delta \hat h(t)|\phi_j^0\ra\,,
\eea
where $\hat P=1-\sum_{l=1}^{M}|\phi_l^0\ra\la\phi_l^0|$.
The perturbed densities and local interaction potentials read
\bea
\delta \rho_{ij}[\delta\mathbf C,\delta\mathbf C^*]
&=&\langle\mathbf{C}^0|\hat a_i^{\dagger}\hat a_j|\delta\mathbf C\rangle
+\langle\delta\mathbf C|\hat a_i^{\dagger}\hat a_j|\mathbf{C}^0\rangle\,,\non
\delta\rho_{isjl}[\delta\mathbf C,\delta\mathbf C^*]
&=&\left[\langle\mathbf{C}^0|\hat a_i^{\dagger}\hat a_s^{\dagger}
\hat a_j\hat a_l|\delta\mathbf C\rangle+\langle\delta\mathbf{C}|
\hat a_i^{\dagger}\hat a_s^{\dagger}\hat a_j\hat a_l|\mathbf C^0\rangle \right]\,,\non
\delta\hat W_{sl}[\delta\phi_l,\delta\phi_s^*]
&=&\int d\mr'\phi_s^{0,*}(\mr')\hat W(\mr-\mr')\delta\phi_l(\mr')
+\int d\mr' \delta\phi_s^{*}(\mr')\hat W(\mr-\mr')\phi_l^0(\mr')\,.
\eea
We make the following ansatz for the perturbed orbitals
\be\label{eq:ansatz_uv_o}
\delta\phi_i(\mathbf{r},t)=
u_i(\mathbf{r}) e^{-i\omega t} +v_i^{*}(\mathbf{r}) e^{i\omega t}\,,
\ee
and for the perturbed coefficients
\be\label{eq:ansatz_uv_c}
\delta \mC=\Cu e^{-i\omega t} +\mathbf{C^{*}_v} e^{i\omega t}\,,
\ee
where $\omega$ is the driving frequency of the external perturbation.
Inserted into Eq.~\eqref{eq:lr_orb} and equating 
the 
same powers of $e^{\mp i\omega t}$ we obtain
\bea\label{eq:lr_orb1}
&&\hat P\Biggl\{\sum_{j=1}^M\left(\hat Z_{ij}^0-
\mu_{ij}^0\right)|u_j\ra+
\sum_{s,j,l=1}^M\rho_{isjl}^0\delta\hat W_{sl}[u_l,v_s]+
\sum_{j=1}^M\delta \rho_{ij}[\Cu,\Cv]\hat h\non
&& +\sum_{s,j,l=1}^M\delta\rho_{isjl}[\Cu,\Cv]\hat W_{sl}^0\Biggr\}|\phi_j^0\ra
-\sum_{j=1}^M\omega\rho_{ij}^0|u_j\ra =-\hat P\sum_{j=1}^M\rho_{ij}^0 f^+|\phi_j^0\ra\,,
\eea
and
\bea\label{eq:lr_orb2}
&&-\hat P^*\Biggl\{\sum_{j=1}^M\left(\hat Z_{ji}^{0}
-\mu_{ij}^{0,*}\right)|v_j\ra+
\sum_{s,j,l=1}^M\rho_{ljsi}^{0}\delta\hat W_{ls}[u_s,v_l]
+\sum_{j=1}^M\delta \rho_{ji}[\Cu,\Cv]\hat h\non
&& +\sum_{s,j,l=1}^M\delta\rho_{ljsi}[\Cu,\Cv]\hat W_{ls}^0\Biggr\}|\phi_j^{0,*}\ra
-\sum_{j=1}^M\omega\rho_{ji}^0|v_j\ra
=\hat P^*\sum_{j=1}^M\rho_{ji}^0 f^-|\phi_j^{0,*}\ra\,.
\eea

\subsubsection{Coefficients part\label{sec:ci}}

In zeroth order we obtain the coefficients' part of 
the MCHB equations for the stationary state
\be
\mathcal{H}^0\mathbf{C}^0=\mathcal{E}^0\mathbf{C}^0\,.
\ee
In first order we get
\be\label{eq:lr_coeff}
i\frac{\partial \delta\mathbf{C}}{\partial t}
=(\mathcal{H}^0-\mathcal{E}^0)\delta\mathbf{C}+\delta\mathcal{H}\mathbf{C}^0\,,
\ee
where the source term is given by
\be
\delta\mathcal{H}=\sum_{k,q}\delta h_{kq}
[\delta\phi_q,\delta\phi_k^*,\delta \hat h]\hat a_k^{\dagger}\hat a_q
+\frac{1}{2}\sum_{k,s,q,l}\hat a_k^{\dagger}\hat a_s^{\dagger}\hat a_l\hat a_q 
\delta W_{ksql}[\delta\phi_q,\delta\phi_l,\delta\phi_k^*,\delta\phi_s^*]\,.
\ee
We introduced the perturbation of the one-body matrix elements
\be
\delta h_{kq}[\delta\phi_q,\delta\phi_k^*,\delta \hat h]
=\int d\mr\phi_k^{0,*}(\mr)\hat h\delta\phi_q(\mr)
+\int d\mr\delta\phi_k^*(\mr)\hat h\phi_q^0(\mr)+\delta h_{kq}^0[\delta \hat h]\,.
\ee
Note that $\delta h_{kq}^0[\delta \hat h]
=\int d\mr\phi_k^{0,*}(\mr)\delta \hat h\phi_q^0(\mr)$ 
contains the perturbation of the potential.
Furthermore, 
the matrix elements of the two-body elements are
\bea
\delta W_{ksql}[\delta\phi_q,\delta\phi_l,\delta\phi_k^*,\delta\phi_s^*]
&=&\iint d\mr d\mr'
\phi_k^{0,*}(\mr)\phi_s^{0,*}(\mr')\hat W(\mr-\mr')\delta\phi_q(\mr)\phi_l^0(\mr')\non
&+&\iint d\mr d\mr'
\phi_k^{0,*}(\mr)\phi_s^{0,*}(\mr')\hat W(\mr-\mr')\phi_q^0(\mr)\delta\phi_l(\mr')\non
&+&\iint d\mr d\mr'
\delta\phi_k^*(\mr)\phi_s^{0,*}(\mr')\hat W(\mr-\mr')\phi_q^0(\mr)\phi_l^0(\mr')\non
&+&\iint d\mr d\mr'
\phi_k^{0,*}(\mr)\delta\phi_s^*(\mr')\hat W(\mr-\mr')\phi_q^0(\mr)\phi_l^0(\mr')\,.
\eea
Inserting the ansatz in Eqs.~\eqref{eq:ansatz_uv_o} 
and \eqref{eq:ansatz_uv_c} into Eq.~\eqref{eq:lr_coeff} 
and equating like powers of $e^{\pm i \omega t}$
we obtain
\bea\label{eq:lr_coeff_om_u}
\omega\Cu&=&(\mathcal{H}^0-\mathcal{E}^0)\Cu \nonumber \\
&+&\left[\sum_{k,q}\delta h_{kq}[u_q,v_k,f^+]\hat a_k^{\dagger}\hat a_q
+\frac{1}{2}\sum_{k,s,q,l}\hat a_k^{\dagger}\hat a_s^{\dagger}\hat a_l\hat a_q
\delta W_{ksql}[u_q,u_l,v_k,v_s]\right]\mathbf{C}^0\,, \
\eea
and
\bea\label{eq:lr_coeff_om_v}
\omega\Cv&=&-(\mathcal{H}^{0,*}-\mathcal{E}^0)\Cv \nonumber \\
&-&\left[\sum_{k,q}\delta h_{qk}[u_k,v_q,f^{-}](\hat a_k^{\dagger}\hat a_q)^*
+\frac{1}{2}\sum_{k,s,q,l}(\hat a_k^{\dagger}\hat a_s^{\dagger}\hat a_l\hat a_q)^*
\delta W_{lqks}[u_k,u_s,v_q,v_l]\right]\mathbf{C}^{0,*}\,, \
\eea
where 
the following notation for the action
of a combination of creation and annihilation operators $\hat O$
on the vector of coefficients is used: 
$\hat O^\ast \mathbf{C}^{0,*} \equiv \{\hat O \mathbf{C}^{0}\}^\ast$.
Summarizing,
Eqs.~(\ref{eq:lr_orb1},\ref{eq:lr_orb2},\ref{eq:lr_coeff_om_u},\ref{eq:lr_coeff_om_v})
are the set of linear-response
equations for a trapped Bose
system interacting via a general two-body
potential $\hat W(\mr-\mr')$.
We note that the above formulation
holds for a symmetric potential $\hat W(\mr,\mr')$ as well.

\subsubsection{Special case: Delta-potential\label{sec:del}}

We now specify the two-body interactions to have the form 
of the widely employed 
delta potential $W(\mr-\mr')=\lambda_0\delta(\mr-\mr')$ \cite{leggett:01}. 
The two-body interaction operators then simplify and read
\be
\hat W_{sl}(\mr)=\lambda_0\phi_s^{*}(\mr)\phi_l(\mr)\,.
\ee
From this we get for the MCTDHB orbital operator [Eq.~(\ref{eq:ham_1st})]
\bea\label{eq:ham_2}
\hat Z_{ij}=\rho_{ij}\hat h+\lambda_0\sum_{s,l=1}^M\rho_{isjl}\phi_s^{*}\phi_l\,,
\eea
and the perturbed two-body interaction operators
\be
\delta\hat W_{sl}[\delta\phi_l,\delta\phi_s^*]
=\lambda_0\left[\phi_s^{0,*}(\mr)\delta\phi_l(\mr)
+\delta\phi_s^{*}(\mr)\phi_l^0(\mr)\right]\,.
\ee
We explicitly write out all expressions 
containing the orbitals' and coefficients' responses $\mus, \mvs,\Cu, \Cv$. 
For the orbitals we get
\bea\label{eq:lr_orb_u_del}
&&\hat P\Biggl\{\sum_{j=1}^M\left(\hat{Z}_{ij}^0
-\mu^0_{ij}\right)|u_j\ra
+\sum_{s,j,l=1}^M\lambda_0
\left[\rho_{islj}^0\phi_s^{0,*}|u_j\ra+\rho_{ijls}^0\phi_s^0 |v_j\ra\right]\non
&&+\sum_{l=1}^M \left[\langle\mathbf{C}^0|\hat a_i^{\dagger}\hat a_l|\Cu\rangle
+\langle\mathbf{C}^{0,*}|(\hat a_i^{\dagger}\hat a_l)^T|\Cv\ra\right]
\hat h\non
&&+\lambda_0\sum_{s,j,l=1}^M
\left[\langle\mathbf{C}^0|\hat a_i^{\dagger}\hat a_s^{\dagger}
\hat a_l\hat a_j|\Cu \rangle+\langle\mathbf C^{0,*}|\left(\hat a_i^{\dagger}
\hat a_s^{\dagger}\hat a_l\hat a_j\right)^T|\Cv\rangle \right]
\phi_s^{0,*}\phi_j^0 \Biggr\}\phi_l^0\non
&& -\sum_{j=1}^M\omega\rho_{ij}^0|u_j\ra =
-\hat P\sum_{j=1}^M\rho_{ij}^0 f^+|\phi_j^0\ra\,,
\eea
and
\bea\label{eq:lr_orb_v_del}
&&-\hat P^*\Biggl\{\sum_{j=1}^M\left(\hat{Z}_{ij}^{0,*}
-\mu^{0,*}_{ij}\right)|v_j\ra
+\sum_{s,j,l=1}^M\lambda_0
\left[\rho_{islj}^{0,*}\phi_s^{0}|v_j\ra
+\rho_{ijls}^{0,*}\phi_s^{0,*}|u_j\ra\right]\non
&&+\sum_{l=1}^M \left[\langle\mathbf{C}^{0,*}|(\hat a_i^{\dagger}\hat a_l)^*|\Cv\rangle
+\langle\mathbf{C}^{0}|(\hat a_i^{\dagger}\hat a_l)^{\dagger}|\Cu\ra\right]
\hat h\non
&&+\lambda_0\sum_{s,j,l=1}^M
\left[\langle\mathbf{C}^{0,*}|(\hat a_i^{\dagger}\hat a_s^{\dagger}
\hat a_l\hat a_j)^*|\Cv \rangle+\langle\mathbf C^{0}|
\left(\hat a_i^{\dagger}\hat a_s^{\dagger}\hat a_l
\hat a_j\right)^{\dagger}|\Cu\rangle
\right]\phi_s^{0}\phi_j^{0,*} \Biggr\}\phi_l^{0,*}\non
&& -\sum_{j=1}^M\omega\rho_{ji}^0|v_j\ra 
=\hat P^*\sum_{j=1}^M\rho_{ji}^0 f^-|\phi_j^{0,*}\ra\,,
\eea
with the notation for operation on the state vector
$\langle\mathbf{C}^{0,*}|\hat O^T \equiv 
\{\hat O |\mathbf C^{0}\rangle\}^T$ and
$\langle\mathbf{C}^{0,*}|\hat O^\ast \equiv 
\{\hat O^\dag |\mathbf C^{0}\rangle\}^T$,
where $T$ 
stands for transpose.
For the coefficients we find
\bea\label{eq:lr_coeff_u_del}
\omega\Cu &=& (\mathcal{H}^0-\mathcal{E}^0)\Cu
+\Bigl[\sum_{k,q}\left(\int d\mr\phi_k^{0,*}\hat h u_q 
+\int d\mr v_k\hat h\phi_q^0 +\delta h_{kq}^0[f^+]\right)\hat a_k^{\dagger}\hat a_q\non
&&+\lambda_0\sum_{k,s,q,l}\hat a_k^{\dagger}\hat a_s^{\dagger}
\hat a_l\hat a_q \left(\int d\mr \phi_k^{0,*}\phi_s^{0,*}
\phi_l^0 u_q+\int d\mr \phi_s^{0,*}\phi_q^0\phi_l^0v_k\right)\Bigr]\mathbf{C}^0\,,
\eea
and
\bea\label{eq:lr_coeff_v_del}
\omega\Cv &=& -(\mathcal{H}^{0,*}-\mathcal{E}^0)\Cv
-\Bigl[\sum_{k,q}\left(\int d\mr\phi_q^{0,*}\hat h u_k 
+\int d\mr v_q\hat h\phi_k^0+\delta h_{qk}^0[f^{-}]\right)
(\hat a_k^{\dagger}\hat a_q)^*\non
&&+\lambda_0\sum_{k,s,q,l}(\hat a_k^{\dagger}\hat a_s^{\dagger}\hat a_l\hat a_q)^*
\left(\int d\mr  \phi_s^{0}\phi_q^{0,*}\phi_l^{0,*}u_k
+\int d\mr \phi_k^{0}\phi_s^{0}\phi_l^{0,*}
v_q\right)\Bigr]\mathbf{C}^{0,*}\,.
\eea
The next step is to cast the linear-response 
Eqs.~\eqref{eq:lr_orb_u_del}, \eqref{eq:lr_orb_v_del}, 
\eqref{eq:lr_coeff_u_del}, and \eqref{eq:lr_coeff_v_del} in matrix form,
which is done in the following subsection.
 
\subsubsection{The linear-response system: 
Matrix form and properties\label{sec:lr_sys_mat}}

The linear-response matrix will act on 
the $2(M+N_{conf})$-dimensional 
vector composed of the orbitals' and coefficients' response 
amplitudes, $\left( \mus,\mvs,\Cu,\Cv \right)^T$\,.
$N_{conf}$ is the number of configurations.
Thus, the general structure of the linear-response matrix, i.e., 
the sizes 
of its submatrices is 
\be
\mL\sim 2\left(\begin{array}{c||c}
M\times M& M\times N_{conf} 
\\\hline\hline N_{conf}\times M& N_{conf}\times N_{conf} \end{array}\right)\,.
\ee
$\mL$ is given explicitly as
\be
\left(\begin{array}{c||c}
\mP^o\mL_{o}& \mP^o \mL_{oc} \\\hline\hline \mL_{co}& \begin{array}{cc}
\mathcal{H}^0-\mathcal{E}^0& 0_c \\ 
0_c& -(\mathcal{H}^{0,*}-\mathcal{E}^0) \end{array} \end{array}\right)\,,
\ee
where $0_c$ is the zero matrix in an $N_{conf}$-dimensional space.
$\mP^o$ is a projector matrix 
that will be discussed later on.
The upper diagonal block $\mL_{o}$ is the purely orbital part. 
The linear response of mean-field fragmented states, 
LR-BMF \cite{grond:12}, 
is a special case of it,
when the state vector of the many-boson system
is represented by a single permanent. 
In the lower diagonal 
we have the (already linear -- in terms of the coefficients) 
CI matrix $\mathcal{H}^0$ and its negative conjugate $-\mathcal{H}^{0,*}$.
The hereafter obtained excitations 
measure the energy relative to the energy of the stationary state $\mathcal{E}^0$. 
The off-diagonals account for the coupling 
between the orbitals' and the coefficients' response. 

In the following we discuss all the involved submatrices.
For the orbital part we have
\be
\mL_{o}=\left(\begin{array}{c|c}
\hat{Z}_{ij}^0-\mu^0_{ij}+\lambda_0\rho_{islj}^0\phi_s^{0,*}\phi_l^0&
  \lambda_0\rho_{ijls}^0\phi_s^{0}\phi_l^0\\
\hline\\
-\lambda_0\rho_{ijls}^{0,*}\phi_s^{0,*}\phi_l^{0,*}&-\left(\hat{Z}_{ij}^{0,*}
-\mu^{0,*}_{ij}\right)-\lambda_0\rho_{islj}^{0,*}\phi_s^{0}\phi_l^{0,*}
\end{array}\right)\,,
\ee
where 
summation over doubled indices
is implicitly assumed. 
The lines separate the blocks for 
$\mus$ and $\mvs$ ($\Cu$ and $\Cv$ below), 
and the coupling between them. 
The upper right submatrix is given as
\be\label{eq:Loc}
\mL_{oc}=
\left(\begin{array}{c|c}
 \hat h \phi_k^0 \langle\mathbf{C}^0_{\vec{n}_i^k}|\cdot 
+ \lambda_0\phi_s^{0,*}\phi_j^0\phi_k^0\langle\mathbf{C}^0_{\vec{n}_{is}^{kj}}|\cdot 
 & \hat h \phi_k^0\langle\mathbf{C}^{0,*}_{\vec{n}_k^i}|\cdot 
+ \lambda_0\phi_s^{0,*}
\phi_j^0\phi_k^0\langle\mathbf{C}^{0,*}_{\vec{n}_{kj}^{is}}|\cdot\\
 \hline
   -\hat h \phi_k^{0,*} \langle\mathbf{C}^0_{\vec{n}_k^i}|\cdot 
- \lambda_0\phi_s^0\phi_j^{0,*}\phi_k^{0,*}
\langle\mathbf{C}^0_{\vec{n}_{kj}^{is}}|\cdot &
    -\hat h \phi_k^{0,*}\langle\mathbf{C}^{0,*}_{\vec{n}_i^k}|\cdot 
- \lambda_0\phi_s^{0}\phi_j^{0,*}\phi_k^{0,*}
\langle\mathbf{C}^{0,*}_{\vec{n}_{is}^{kj}}|\cdot 
 \end{array} \right)\,.
\ee
Each block is an $M\times N_{conf}$ matrix, i.e., 
all indices are summed over except for $i$, 
which accounts for having $M$ lines in each block. 
We further use in the matrix
representation of the linear-response system the 
compact notation
$|\mC^0_{\vec{n}_i^k}\ra \equiv \hat a_k^{\dagger} \hat a_i|\mC^0\ra$ 
and
$|\mC^0_{\vec{n}_{is}^{kj}}\ra \equiv \hat a_j^{\dagger} 
\hat a_k^{\dagger} \hat a_s \hat a_i|\mC^0\ra$.
The lower left submatrix is given as
\bea\label{eq:Lco}
&&\mL_{co}=\\
&& \left(\begin{array}{c|c}
 |\mC^0_{\vec{n}_i^k}\ra \int
\phi_k^{0,*}\hat h\cdot
+\lambda_0|\mC^0_{\vec{n}_{il}^{ks}}\ra \int
\phi_k^{0,*}\phi_s^{0,*}\phi_l^0\cdot &  
 |\mC^0_{\vec{n}_k^i}\ra \int
\phi_k^0\hat h\cdot
+\lambda_0|\mC^0_{\vec{n}_{kl}^{is}}\ra \int
\phi_s^{0,*}\phi_k^{0}\phi_l^0\cdot\\
 \hline
- |\mC^{0*}_{\vec{n}_k^i}\ra\int
\phi_k^{0,*}\hat h\cdot 
-\lambda_0|\mC^{0,*}_{\vec{n}_{kl}^{is}}\ra 
\int
\phi_s^{0}\phi_k^{0,*}\phi_l^{0,*}\cdot&  
-|\mC^{0*}_{\vec{n}_i^k}\ra\int
\phi_k^0\hat h\cdot
-\lambda_0|\mC^{0,*}_{\vec{n}_{il}^{ks}}\ra\cdot 
\int
\phi_k^{0}\phi_s^{0}\phi_l^{0,*}\cdot
 \end{array} \right)\,.\nonu
\eea
Here, for each $i$, an orbital (one-body function) is `lying' in each block. 
The dot signifies that a scalar product has to be taken with the 
corresponding vector element 
where the linear-response 
matrix acts on. 
For $\mL_{oc}$ it is the Euclidean scalar product 
in an $N_{conf}$-dimensional space. 
For $\mL_{co}$ it means integration over space.

In order to find the orthonormalization relations, 
we analyze the symmetry of the response matrix $\mL$. 
This can be achieved by generalizing the discussion of 
linear-response matrices of GP \cite{castin:98} and BMF \cite{grond:12}.  
The orbitals' sub-diagonal $\mL_o$ can be analyzed in analogy 
to the linear-response matrix of BMF. 
First, one finds 
a time-reversal spin-flip-like symmetry
\be\label{eq:sigma1_1}
\boldsymbol{\Sigma_1}^o\mL_o\boldsymbol{\Sigma_1}^o=-\left(\mL_o\right)^*\,,
\ee
where the matrix $[\Sigma_1]_{ij}=\delta_{i,j-M}+\delta_{i-M,j}$ 
($i,j=1,...,2M$) permutes the $i$-th and the $M+i$-th 
rows, 
just as the first Pauli matrix 
$\sigma_1=\bigl(\begin{smallmatrix}0&1\\1&0\end{smallmatrix}\bigr)$ does for $M=1$. 
Further, we have
\be\label{eq:sigma3_1}
\boldsymbol{\Sigma_3}^o\mL_o\boldsymbol{\Sigma_3}^o=\left(\mL_o\right)^{\dagger}\,,
\ee
with the matrix 
\be
[\Sigma_3^o]_{ij}=\Biggl\{\begin{array}{ll}\delta_{i,j}&,
\quad\mathrm{for}\quad i,j\le M\\ -\delta_{i,j}&,
\quad\mathrm{for}\quad i,j> M \end{array}\;.
\ee
For the case $M=1$, 
$\boldsymbol{\Sigma_3}^o$ boils down 
to the third Pauli matrix $\sigma_3=\bigl(\begin{smallmatrix}1&0\\0&-1\end{smallmatrix}\bigr)$.
From Eq.~\eqref{eq:sigma1_1} we learn that, 
whenever $(|\mathbf u^k\ra,|\mathbf v^k\ra)^T$ is an eigenvector of $\mL_o$ 
with eigenvalue $\omega_k$, 
then $(|\mathbf v^{k,*}\ra,|\mathbf u^{k,*}\ra)^T$ is 
an eigenvector with eigenvalue $-(\omega_{k})^*$. 
From Eq.~\eqref{eq:sigma3_1} we find 
that $(|\mathbf u^k\ra,-|\mathbf v^k\ra)^T$ is 
an eigenvector of $\left(\mL_o\right)^{\dagger}$ 
with eigenvalue $\omega_k$, 
which allows us to construct the adjoint basis. 

The same symmetries hold for the lower diagonal submatrix 
of $\mL$ with respect to the matrices
\be\label{eq:sigma1_2}
\boldsymbol{\Sigma_1}^c=\left(\begin{array}{c c} 0_c & 1_c 
\\ 1_c & 0_c\end{array}\right),\quad\boldsymbol{\Sigma_3}^c
=\left(\begin{array}{c c} 1_c & 0_c \\ 0_c & -1_c\end{array}\right)\,,
\ee
where $1_c$ and $0_c$ are the unit and zero matrices 
in an $N_{conf}$-dimensional space, respectively. 
For the off-diagonal submatrices of $\mL$ we find that the diagonal 
blocks 
of $\mL_{oc}$ are the hermitian conjugates of the 
diagonals of $\mL_{co}$, 
while the upper (lower) off-diagonal of $\mL_{oc}$ is the negative 
hermitian conjugate of the lower (upper) 
off-diagonal of $\mL_{co}$. 
Hence, the full linear-response matrix $\mL$ obeys the same 
symmetries with respect to matrices composed of the symmetry 
matrices of the orbitals' and coefficients' parts
\be\label{eq:sigma1_3}
\boldsymbol{\Sigma_1}=\left(\begin{array}{c c} 
\boldsymbol{\Sigma_1}^o & 0_{oc} \\ 0_{co} & 
\boldsymbol{\Sigma_1}^c\end{array}\right),\quad\boldsymbol{\Sigma_3}
=\left(\begin{array}{c c} \boldsymbol{\Sigma_3}^o & 0_{oc} 
\\ 0_{co} & \boldsymbol{\Sigma_3}^{c}\end{array}\right)\,.
\ee
Here, we used the zero matrices $0_{co}$ and $0_{oc}$ with 
the size of the corresponding non-square submatrices. 
Finally, from those symmetries we obtain the following 
orthonormality condition for the response amplitudes of the 
orbitals and coefficients (with excitation index $k$)
\bea\label{eq:orth_1}
&&\la \mathbf u^k|\mathbf u^{k'}\ra-\la \mathbf v^k|\mathbf v^{k'}\ra
+\la \Cu^k|\Cu^{k'}\ra-\la \Cv^k|\Cv^{k'}\ra=\delta_{kk'}\,,\non
&&\la \mathbf v^k|\mathbf u^{k',*}\ra-\la \mathbf u^k|\mathbf v^{k',*}\ra
+\la \Cv^k|\Cu^{k',*}\ra-\la \Cu^k|\Cv^{k',*}\ra=0\,.
\eea
From this we immediately see that an excitation 
can be either orbital-like or coefficients-like, 
or a mixture of both, 
see for more details Sec.~\ref{sec:type} below.

The linear-response equations read
\bea\label{eq:lr}
&&\left(\mP\mL-\mM\omega\right)\left(\begin{array}{c}|\mathbf{u}\ra 
\\ |\mathbf{v}\ra\\|\Cu\ra\\|\Cv\ra\end{array}\right)
=\mM\boldsymbol{\mathcal{P}}\left(\begin{array}{c}-f^+|\boldsymbol{\phi^0}\ra 
\\ f^{-}|\boldsymbol{\phi^{0,*}}\ra\\ 
-\delta h_{kq}^0[f^+]\hat a_k^{\dagger}\hat a_q |\mC^0\ra\\\delta h_{qk}^0[f^{-}]
\left(\hat a_k^{\dagger}\hat a_q\right)^* |\mC^{0,*}\ra\end{array}\right)\,,
\eea
where we again sum over double indices. 
The projector matrix contains twice as many projectors 
as the number of orbitals $M$ ($i,j=1,...,2M$)
\be
\mathcal{P}_{ij}^o=\left\{\begin{array}{l} 
\hat P\,,\quad\mathrm{for}\quad i=j\le M \\ 
\hat P^*\,,\quad\mathrm{for}\quad i=j> M \\ 0\,,\quad(i\neq j)\end{array}\right.\;,
\ee
leading to the full matrix
\be
\mP=\left(\begin{array}{c c} \mP^o & 0_{oc}\\0_{co}&1_c\end{array}\right)\,.
\ee
Since the projector appears on both sides of Eq.~\eqref{eq:lr}, 
but not on the term proportional to the driving frequency $\omega$, 
the solution $\left( \mus,\mvs,\Cu,\Cv \right)^T$ 
must be orthogonal to all the ground-state orbitals. 
We therefore can safely replace $\mP\mL\rightarrow\mP\mL\mP$ in Eq.~\eqref{eq:lr}, 
in order to guarantee the above-discussed symmetries also in presence of the projector.
Moreover, in the linear-response equation we have a metric 
\be
\mM=\left(\begin{array}{cccc} \mrh^0&&&\\&\mrh^{0,*}&&
\\&&1_c&\\&&&1_c \end{array}\right)\,
\ee
in which the stationary reduced one-particle density matrix
$\mrh^0 = \{\rho^0_{ij}\}$ 
appears.
The other elements are zero. 
In order to render the left-hand side 
of Eq.~\eqref{eq:lr} an eigenvalue problem, 
and at the same time preserve the symmetries of the corresponding matrix, 
we first take 
the square root 
of the metric $\mM$ 
and cast the linear-response equation into the form
\bea\label{eq:lr2}
&&\left(\overline{\mL}-\omega\right)\left(\begin{array}{c}|\overline{\mathbf{u}}\ra 
\\ |\overline{\mathbf{v}}\ra\\|\overline{\Cu}\ra\\|\overline{\Cv}\ra\end{array}\right)
=\left(\begin{array}{c}-(\mrh^0)^{1/2}\mP^of^+|\boldsymbol{\phi^0}\ra 
\\ (\mrh^{0,*})^{1/2}\mP^{o,*} f^{-}|\boldsymbol{\phi^{0,*}}\ra\\ 
-\delta h_{kq}^0[f^+]\hat a_k^{\dagger}\hat a_q |\mC^0\ra\\\delta h_{qk}^0[f^{-}]
\left(\hat a_k^{\dagger}\hat a_q\right)^* |\mC^{0,*}\ra\end{array}\right)\,,
\eea
where 
the final form of the linear-response matrix reads
\be\label{eq:eigL}
\overline{\mL}=\mM^{-1/2}\mP\mL\mP\mM^{-1/2}\,.
\ee
Since $\mM$ and $\mM^{-1/2}$ have the same symmetry properties as $\mL$, 
this is true also for $\overline{\mL}$. 
Furthermore, the final form of the response amplitudes is 
\be\label{eq:uup}
\left(\begin{array}{c}|\overline{\mus}\ra\\
|\overline{\mvs}\ra\\|\overline{\Cu}\ra\\|\overline{\Cv}\ra\end{array}\right)
=\mM^{1/2}\left(\begin{array}{c}|\mus\ra
\\|\mvs\ra\\|\Cu\ra\\|\Cv\ra\end{array}\right)
=\left(\begin{array}{c}(\mrh^0)^{1/2}|\mus\ra\\(\mrh^{0,*})^{1/2}|\mvs\ra
\\|\Cu\ra\\|\Cv\ra\end{array}\right)\,.
\ee
We find that 
the same orthonormalization relations as in Eq.~\eqref{eq:orth_1},
\bea\label{eq:orth_2}
&&\la \overline{\mathbf u}^k|\overline{\mathbf u}^{k'}\ra
-\la \overline{\mathbf v}^k|\overline{\mathbf v}^{k'}\ra
+\la \overline{\Cu}^k|\overline{\Cu}^{k'}\ra
-\la \overline{\Cv}^k|\overline{\Cv}^{k'}\ra=\delta_{kk'}\,,\non
&&\la \overline{\mathbf v}^k|\overline{\mathbf u}^{k',*}\ra
-\la \overline{\mathbf u}^k|\overline{\mathbf v}^{k',*}\ra
+\la \overline{\Cv}^k|\overline{\Cu}^{k',*}\ra
-\la \overline{\Cu}^k|\overline{\Cv}^{k',*}\ra=0\,,
\eea
are satisfied.
For convenience of presentation,
we from now on omit
the `bar' over the linear-response quantities.

In order to find the excitation energies $\omega_k$ 
in Eq.~\eqref{eq:lr2} one has to solve the eigenvalue problem
\be\label{eq:evL}
\boldsymbol{\mathcal{P}}\boldsymbol{\mathcal{L}}
\boldsymbol{\mathcal{P}}\left(\begin{array}{c}|\mathbf{u}^k\ra
\\|\mathbf{v}^k\ra\\|\Cu^k\ra\\|\Cv^k\ra\end{array}\right)
=\omega_k\left(\begin{array}{c}|\mathbf{u}^k\ra\\|\mathbf{v}^k\ra
\\|\Cu^k\ra\\|\Cv^k\ra\end{array}\right)\,
\ee
(a redundant $\boldsymbol{\mathcal{P}}$ is added
to both sides of $\boldsymbol{\mathcal{L}}$
to remind that the response is
in the complementary space of the orbitals).
Now we solve Eq.~\eqref{eq:lr2} by expanding the response vectors 
as well as the perturbation 
in the eigenvectors 
of $\boldsymbol{\mathcal{P}}\boldsymbol{\mathcal{L}}\boldsymbol{\mathcal{P}}$ 
orthogonal to the stationary orbitals $\phi_i^0(\mr)$. 
The ansatz for the response amplitudes then reads
\be\label{eq:ansatz1}
\left(\begin{array}{c}|\mathbf u \ra\\ 
|\mathbf v\ra\\|\Cu\ra\\|\Cv\ra\end{array}\right)
=\sum_{k} c_k\left(\begin{array}{c}|\mus^k \ra
\\ |\mvs^k\ra\\|\Cu^k\ra\\|\Cv^k\ra\end{array}\right)\,,
\ee
and for the perturbation
\be\label{eq:ansatz2}
\left(\begin{array}{c}-(\mrh^0)^{1/2}\mP^of^+|\boldsymbol{\phi^0}\ra 
\\ (\mrh^{0,*})^{1/2}\mP^{o,*} f^{-}|\boldsymbol{\phi^{0,*}}\ra
\\ -\delta h_{kq}^0[f^+]\hat a_k^{\dagger}\hat a_q |\mC^0\ra
\\\delta h_{qk}^0[f^{-}]\left(\hat a_k^{\dagger}\hat a_q\right)^* 
|\mC^{0,*}\ra\end{array}\right)
=-\sum_{k} \gamma_k\left(\begin{array}{c}|\mus^k \ra
\\ |\mvs^k\ra\\|\Cu^k\ra\\|\Cv^k\ra\end{array}\right)\,.
\ee
Now $c_k$ and $\gamma_k$ have to be determined. 
Substituting Eqs.~\eqref{eq:ansatz1} and \eqref{eq:ansatz2} 
into Eq.~\eqref{eq:lr2}, 
we obtain
\bea\label{eq:coeff_comp_det}
&&\sum_k c_k(\omega_k-\omega)\left(\begin{array}{c}|\mus^k \ra
\\ |\mvs^k\ra\\|\Cu^k\ra\\|\Cv^k\ra\end{array}\right)
=-\sum_{k} \gamma_k\left(\begin{array}{c}|\mus^k \ra
\\ |\mvs^k\ra\\|\Cu^k\ra\\|\Cv^k\ra\end{array}\right)\,,
\eea
where $\omega_k$ is defined in Eq.~\eqref{eq:evL}. 
From comparing the coefficients in Eq.~\eqref{eq:coeff_comp_det} 
we get an expression for the $c_k$. 
Inserted into Eq.~\eqref{eq:ansatz1} 
leads to a solution 
for the response amplitudes of the form
\be\label{eq:response_ampl}
\left(\begin{array}{c}|\mus \ra
\\ |\mvs\ra\\|\Cu\ra\\|\Cv\ra\end{array}\right)
=\sum_k\frac{\gamma_k}{\omega-\omega_k}\left(\begin{array}{c}|\mus^k \ra
\\ |\mvs^k\ra\\|\Cu^k\ra\\|\Cv^k\ra\end{array}\right)\,.
\ee
Reinserting the amplitudes into the ansatz for the orbitals, 
Eqs.~\eqref{eq:ansatz_uv_o}, \eqref{eq:ansatz_uv_c} and \eqref{eq:uup}, 
we arrive at the final solution for the time-dependent orbitals 
in linear response:
\be\label{eq:response_orb}
\boldsymbol{\phi}(\mathbf{r},t)
=\boldsymbol{\phi^0}(\mathbf{r})
+(\mrh^0)^{-1/2}\sum_k\left[ \gamma_k \mus^k e^{-i\omega t}
+\gamma_k^* \mvs^{k,*} e^{i\omega t}\right]/(\omega-\omega_k)\,,
\ee
where $\boldsymbol{\phi}(\mathbf{r},t) = \{\phi_i(\mathbf{r},t)\}$ and
$\boldsymbol{\phi^0}(\mathbf{r}) = \{\phi^0_i(\mathbf{r})\}$
are column vectors collecting the respective orbitals.
Thus, the orbitals and with them the density show the largest response 
at the frequencies $\omega_k$. 
Moreover, the response for a fixed frequency $\omega_k$ 
is not necessarily equally strong for all the orbitals. 
This is because the components of the 
response amplitudes $u_j^k$ and $v_j^k$ are not normalized, 
but rather the whole amplitude vector [see Eq.~\eqref{eq:orth_2}]. 
The coefficients read to first order
\be\label{eq:response_coeff}
\mathbf{C}(t)=e^{-i\mathcal{E}^0 t}\left\{\mathbf{C^0}+\sum_k
\left[ \gamma_k \Cu^k e^{-i\omega t}+\gamma_k^* \Cv^{k,*} 
e^{i\omega t}\right]/(\omega-\omega_k)\right\}\,.
\ee
The response weights, 
which quantify the intensity of the response, 
are given as
\bea\label{eq:weight}
\gamma_k&=&\la\mus^k|f^+(\mrh^0)^{1/2}|\boldsymbol{\phi^0}\ra
+\la\mvs^k|f^-(\mrh^{0,*})^{1/2}|\boldsymbol{\phi^{0,*}}\ra\\
&&+\left(\int d\mr\phi_i^{0,*}f^+\phi_j^0\right)
\la\Cu^k|\hat a_i^{\dagger} \hat a_j|\mC^0\ra
+\left(\int d\mr\phi_j^{0,*}f^{-}\phi_i^0\right)
\la\Cv^k|\left(\hat a_i^{\dagger} \hat a_j\right)^*|\mC^{0,*}\ra\,.\nonu
\eea
Similar to the 
orbitals [Eq.~\eqref{eq:response_orb}], 
they are dominated by the largest components of the response amplitudes.

\subsection{Density oscillations \label{sec:den}}

In order to be able to calculate the density oscillations, 
we need first the 
reduced one-body density matrix 
\cite{sakmann:08,Lowdin,Yukalov_Book} to first order. 
Its elements are given by
\bea\label{eq:response_rhoij}
\rho_{ij}(t)&=&\la \mC(t)|\hat a_i^{\dagger}\hat a_j|\mC(t)\ra\non
&\approx&\rho^0_{ij}+\sum_k\Bigl[ 
\gamma_k \left(\la \mC^0|\hat a_i^{\dagger}\hat a_j|\Cu^k\ra
+\la \Cv^{k,*}|\hat a_i^{\dagger}\hat a_j|\mC^0\ra\right) e^{-i\omega t}\non
&&+\gamma_k^* \left(\la \mC^0|\hat a_i^{\dagger}\hat a_j|\Cv^{k,*}\ra
+\la \Cu^{k}|\hat a_i^{\dagger}\hat a_j|\mC^0\ra\right) 
e^{i\omega t}\Bigr]/(\omega-\omega_k).
\eea
Together with Eq.~\eqref{eq:response_orb} we obtain
\bea\label{eq:density_response}
& & \rho(\mr,t)=\sum_{i,j=1}^M\rho_{ij}(t)\phi_i^*(\mr,t)\phi_j(\mr,t)  \approx \rho^0(\mr) \\
& & \, + 2\sum_k \frac{1}{\omega-\omega_k} 
\left\{ \Re{[\gamma_k]}\Re{[\Delta\rho^k_o(\mr)+\Delta\rho^k_c(\mr)]} -
      \Im{[\gamma_k]}\Im{[\Delta\rho^k_o(\mr)+\Delta\rho^k_c(\mr)]}\right\}\cos{(\omega t)}
\nonumber \\
& & \, + 2\sum_k \frac{1}{\omega-\omega_k} 
\left\{\Re{[\gamma_k]}\Im{[\Delta\rho^k_o(\mr)+\Delta\rho^k_c(\mr)]} +
  \Im{[\gamma_k]}\Re{[\Delta\rho^k_o(\mr)+\Delta\rho^k_c(\mr)]}\right\}\sin{(\omega t)}. 
\nonumber \
\eea
The density shows the largest response at the linear-response resonance frequencies. 
For simplicity,
we assume real stationary orbitals $\phi_i^0(\mathbf{r})$
and reduced one-body density matrix $\mrh^0$. 
We then obtain for the oscillatory part of the real-space density
the orbitals' contribution
\be\label{eq:density_osc_orb}
\Delta\rho^{k}_o(\mathbf{r})=
\sum_{i,j=1}^M (\mrh^0)^{1/2}_{ij}
\phi_i^0(\mathbf{r})
\left\{u_j^k(\mathbf{r})+v_j^{k}(\mathbf{r}) \right\}\,,
\ee
and the coefficients' contribution
\be\label{eq:density_osc_coef}
\Delta\rho^{k}_c(\mathbf{r})=
\la \mC^0|\hat \rho^0(\mr)|\Cu^k\ra+\la \Cv^{k,*}|\hat \rho^0(\mr)|\mC^0\ra,
\ee
where we defined 
$\hat \rho^0(\mr) = 
\sum_{i,j=1}^{M}\hat a_i^{\dagger}\hat a_j\phi_i^{0,*}(\mr)\phi_j^{0}(\mr)$.

\subsection{Type of excitations\label{sec:type}}

We finally reexamine 
the equation for the norm of the 
LR-MCTDHB response amplitudes, Eq.~\eqref{eq:orth_2}.
The left-hand side of this equation consists of the sum of
the norm of the orbitals' response amplitudes 
and the CI response amplitudes.
Hence, in order for Eq.~\eqref{eq:orth_2} to hold, 
it suffices 
that either the orbitals' 
or the CI part of it is finite (and equal to one).
Besides that, there can exist interesting types of mixed excitations.
Let us characterize the type of the excitation 
by the norm of its CI response amplitude:
\be\label{eq:type}
t_k=\la \Cu^k|\Cu^{k}\ra-\la \Cv^k|\Cv^{k}\ra
    =1-\left(\la \mathbf u^k|\mathbf u^{k}\ra
-\la \mathbf v^k|\mathbf v^{k}\ra\right)\,.
\ee 
Thus, $t_k=0$ implies a purely orbital-like excitation, 
whereas $t_k=1$ indicates a purely CI-like one.
Although, in general, the type of excitation is rather 
dependent on the basis and the number of modes used,
the quantity $t_k$ is useful
to understand the nature of the excitations,
and will be analyzed in the 
following sections.
We note that $t_k$ is not an observable.

\section{Illustrative examples\label{sec:app}}

We now turn to the application of the LR-MCTDHB theory. 
We choose for simplicity the commonly used contact inter-particle interaction potential. 
First, we study the response 
of many-boson systems trapped in an harmonic potential, 
and then turn to a more complex system -- a shallow double-well potential. 
We obtain the excitation spectra by first calculating the ground state of 
the corresponding MCTDHB(M) equations, see Sec.~\ref{sec:mctdhb}. 
Thereafter, we explicitly construct the linear-response matrix, 
Eq.~\eqref{eq:eigL}, and diagonalize it numerically
solving thereby the eigenvalue problem Eq.~\eqref{eq:evL}.
We recall here that the MCTDHB(1) level of theory 
is identical to the famous Gross-Pitaevskii mean field, 
hence the LR-MCTDHB(1) equations
are the familiar LR-GP equations, 
often referred to in literature 
as the particle conserving Bogoliubov-de Gennes 
equations \cite{castin:98}.

\subsection{Bose-Einstein condensates in a
harmonic trap \label{sec:harm}}

We start with the simple example of a BEC in a one-dimensional 
harmonic potential
\be\label{eq:pot_h}
V(x)=\omega_{ho}^2 x^2/2\,,
\ee
with frequency $\omega_{ho}=\sqrt{2}$. 
We study the linear response of systems made of $N=100$  
bosons with inter-particle interaction 
strength $\lambda_0=0.01$ and of $N=10$ bosons with $\lambda_0=0.1$.
The ground states 
of these systems are well described by the GP equation,
since the fragmentation, i.e., the population of higher orbitals, 
is negligible: $0.001\%$ for $N=100$ and $0.01\%$ for $N=10$, respectively.
These systems are characterized by the 
same non-linear parameter $\lambda_0N=1.0$, 
implying that they have the same GP and, hence, 
the same BdG linear-response solutions.
In Sec.~\ref{subsec:h100} we compare the 
linear-response spectra of these systems
obtained within the frameworks of the standard LR-GP (BdG) 
mean-field-based theory and our many-body LR-MCTDHB(2).
Thereafter, in Sec.~\ref{subsec:h10} we address the 
convergence of the LR-MCTDHB($M$) predictions for the system of $N=10$ bosons
by contrasting the spectra computed 
at different levels $M=1,2,4,5$ of the theory.
We recall that in the harmonic potential one can 
separate the center-of-mass (`CM') and relative-motion (`rel') degrees of freedom.
Below, we examine the computed excitations in this context.

\subsubsection{LR-MCTDHB and LR-GP results for N=10 and N=100\label{subsec:h100}}

\begin{figure}[ht]
\vspace{-2.0cm}
\begin{tabular}{ll}
\vspace{-1.5cm}
\includegraphics[width=.55\columnwidth,angle=-90]
{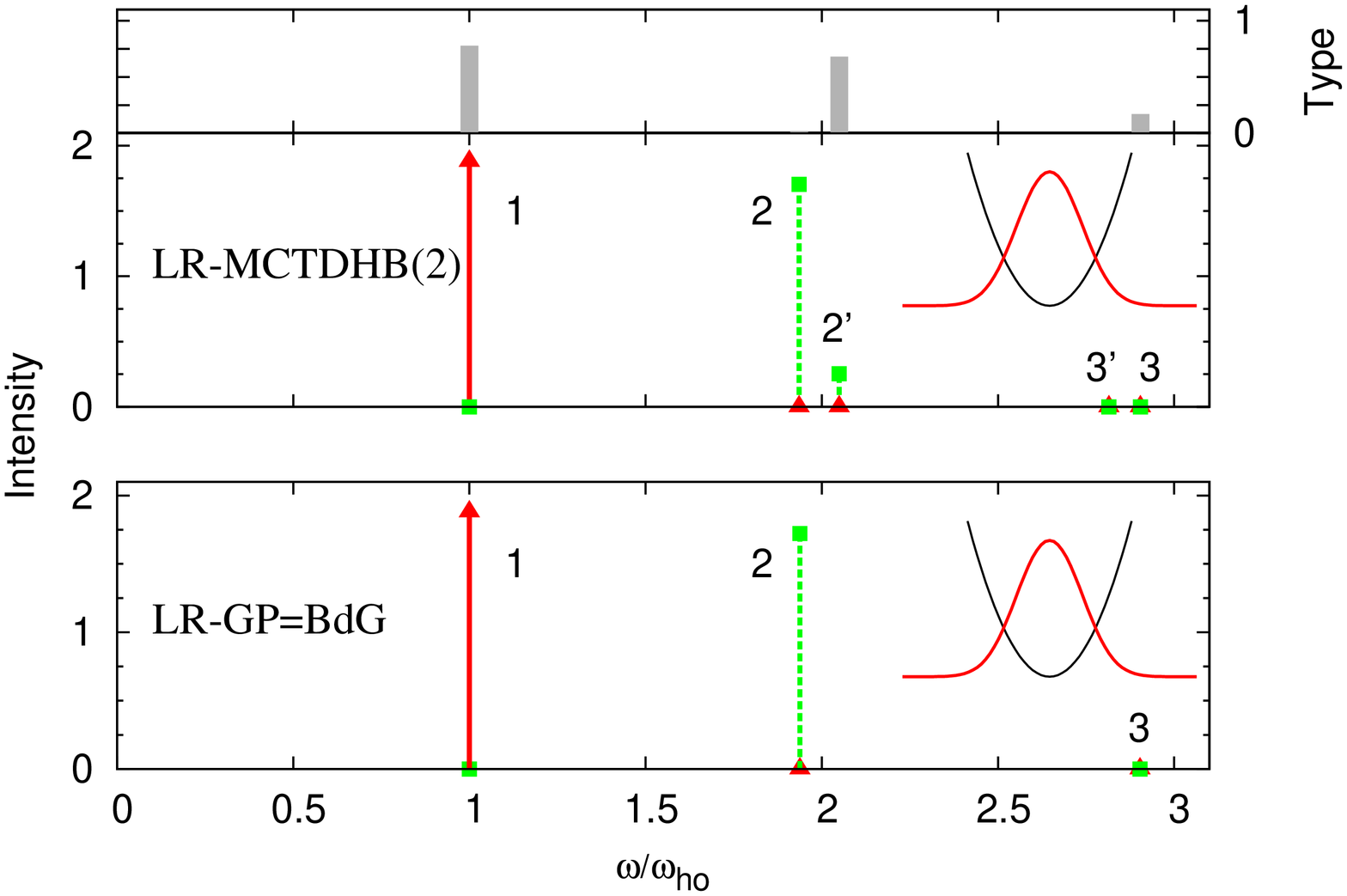}\\           
\vspace{-1cm}
\includegraphics[width=.55\columnwidth,angle=-90]
{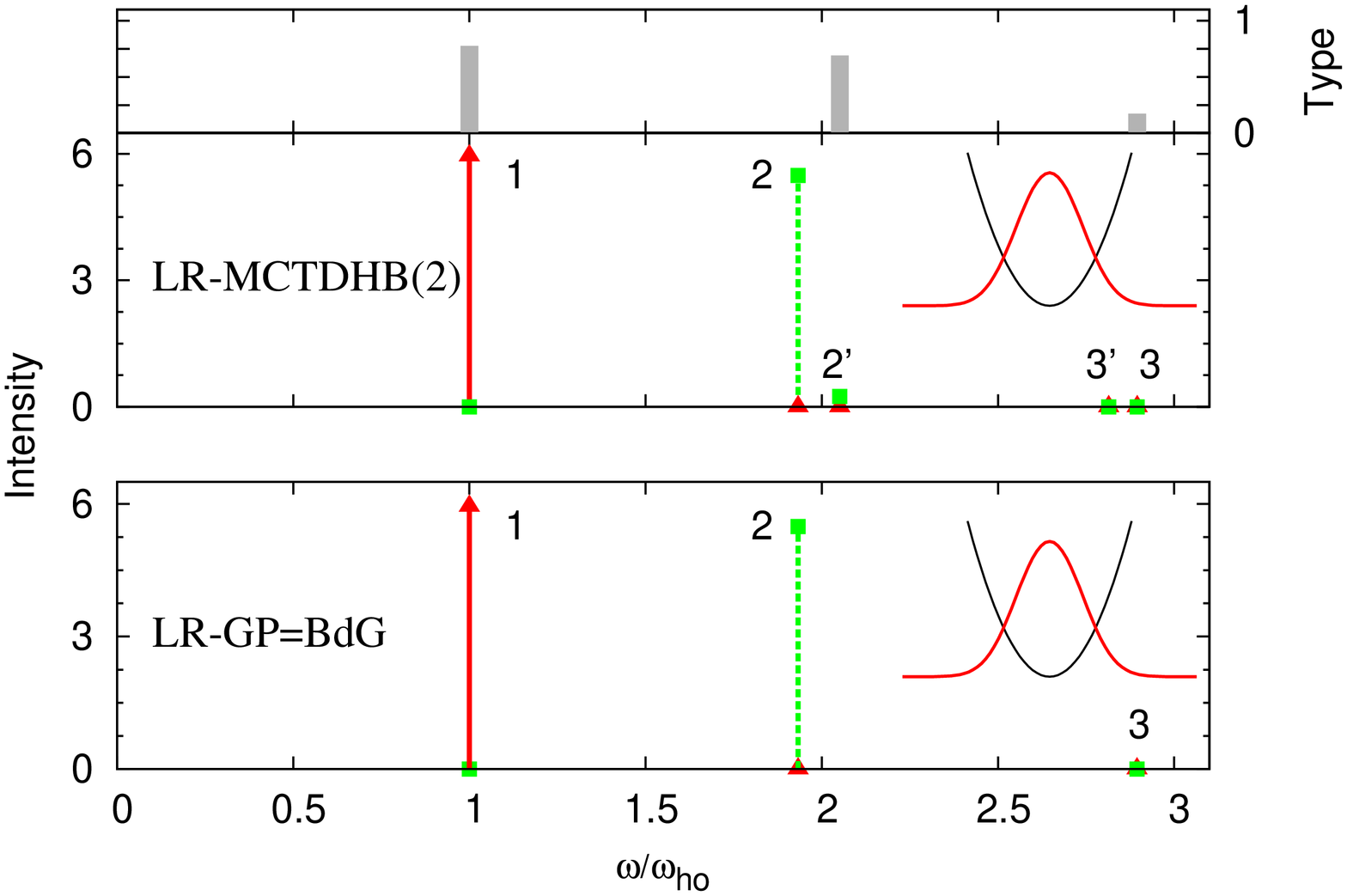}           
\end{tabular}
\vspace{0.5cm}
\caption{(Color online) Comparison 
of the standard LR-GP (BdG) 
and LR-MCTDHB(2) linear-response spectra computed 
for the systems of $N=100$ (lower two panels) 
and $N=10$ (upper two panels) bosons 
trapped in a one-dimensional 
harmonic potential and subject to 
linear and quadratic perturbations. 
The trap potential and ground-state density (red solid line) are shown in the insets.
The lower part of each panel displays 
excitation frequencies on the x-axis and response weights,
Eq.~\eqref{eq:weight},
on the y-axis for linear (odd) $f^+=f^-=x$ and quadratic (even) $f^+=f^-=x^2$ 
perturbations depicted by the solid (red triangle) 
and dashed (green squares) lines (symbols), respectively. 
The mean-field-based excitations, well described by both theories, are labeled 1,2,3.
The many-body excitations, 
not describable by the BdG theory, 
are labeled as 2' and 3' 
(for more details see the text and Fig.~\ref{fig:dens0}). 
The upper parts of the panels display the type of excitation $t_k$
computed as in Eq.~\eqref{eq:type}: 
$t_k=0$ implies a purely orbital-like excitation, 
$t_k=1$, a purely CI one.
All quantities are dimensionless.
\label{fig:h1_N100_N10_ho}}
\end{figure}

In Fig.~\ref{fig:h1_N100_N10_ho} we compare the 
LR-GP and LR-MCTDHB(2) excitation spectra for 
systems of $N=100$ (lower two panels) and $N=10$ 
(upper two panels) bosons trapped in 
the harmonic potential (\ref{eq:pot_h}).
The ground-state densities obtained for these systems  
within the GP and MCTDHB(2) theories are very similar
and schematically shown in the insets together 
with the trapping potential.
The x-axes in Fig.~\ref{fig:h1_N100_N10_ho} 
indicate excitation energies in units of 
the trap frequency $\omega_{ho}$. 
The height of the lines or, equivalently, 
the position of the points along the y-axes, 
indicate response weights $\gamma_k$ 
[see Eq.~\eqref{eq:weight}]. 
We have chosen linear $f^+=f^-=x$ and quadratic $f^+=f^-=x^2$ perturbations
to study both even- and odd-parity excitations separately.
In Fig.~\ref{fig:h1_N100_N10_ho} the solid (red) 
lines with triangles correspond to the odd excitations, 
and the dashed (green) lines with squares to even ones. 

Let us first analyze the excitation spectrum of the $N=10$ system. 
The main observation is that the LR-MCTDHB(2) 
spectrum has more lines then the respective LR-GP one.
Hence, there are low-lying excitations not described by the BdG theory.
On the other hand, the position and response weights 
(intensities) of the most intense lines in the spectra
are well described by BdG 
and essentially 
reproduce the results of the many-body theory.

\begin{figure}[ht]
\vglue -1.0 truecm
\includegraphics[width=.95\columnwidth]
{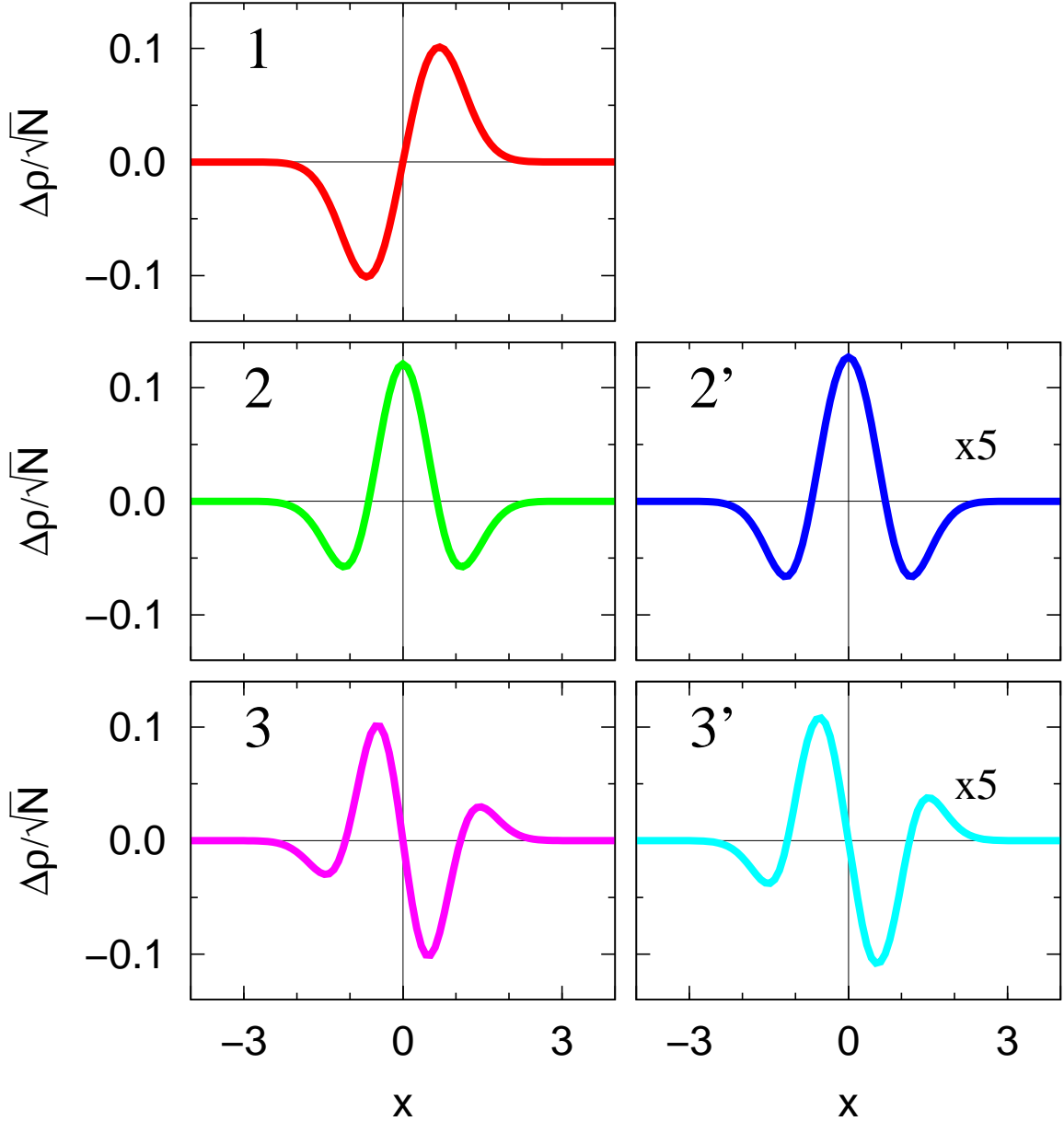}
\caption{(Color online) Density responses for $N=10$
bosons in a one-dimensional harmonic potential.
Plotted are the real parts of the response densities 
$\Delta\rho(x)=\Re[\Delta\rho^k_o(x)+\Delta\rho^k_c(x)]$ 
corresponding to real-valued $\gamma_k$'s, see text for details.
All excitations are numerated according 
to the underlying nodal structures (number of nodes) 
in the respective response density. 
The LR-GP (BdG) results for the first few excitations 
are depicted in the left column 
of the figure.
The LR-MCTDHB results for these excitations are quite similar and, 
therefore, not shown. 
The many-body excitations,
not describable by the BdG theory, 
are labeled by the numbers with a prime, 
and magnified for better visibility.
All quantities are dimensionless.}
\label{fig:dens0}
\end{figure}

To gain more insight into the similarities and 
differences we plot in Fig.~\ref{fig:dens0}
the LR-GP and LR-MCTDHB density responses of the corresponding excitations. 
For their computation we have used the components of the eigenvectors, 
Eq.~\eqref{eq:evL}.
The total density response is a sum of 
the orbital 
[Eq.~\eqref{eq:density_osc_orb}] 
and CI 
[Eq.~\eqref{eq:density_osc_coef}] 
contributions.
From the derivation it is clear that the response weights 
$\gamma_k$ and response amplitudes 
enter Eq.~\eqref{eq:response_ampl} as a product 
and, therefore, there is a degree of freedom to choose the phase of the 
response amplitudes such that the $\gamma_k$ 
are real numbers.
From now on, e.g., in Fig.~\ref{fig:dens0}, 
we plot the real parts of the response densities
$\Delta\rho(x)=\Re[\Delta\rho^k_o(x)+\Delta\rho^k_c(x)]$ 
corresponding 
to real-valued $\gamma_k$.

In harmonic traps the lowest-in-energy excitation 
corresponds to the so-called dipole oscillation 
and has ungerade (odd) symmetry. 
It has a remarkable property -- 
it is a collective excitation of the center-of-mass motion and, 
due to the separability of the center-of-mass and relative motions 
in harmonic traps, it does not depend on the inter-particle interaction.
Hence, its excitation energy must always be $\omega_{ho}$. 
We see that this excitation is nicely described by the LR-GP and LR-MCTDHB theories.
The lowest-in-energy excitation, 
labeled 1 in Fig.~\ref{fig:h1_N100_N10_ho} and Fig.~\ref{fig:dens0}, 
responds
only to the $x$-shaped perturbation, 
while its contribution to the response of the 
$x^2$-perturbation is zero.
In Fig.~\ref{fig:dens0} one can see that the 
density response of the lowest-in-energy excitation has the 
expected one-node profile,
similar to the shape of the first excited orbital 
of a harmonic oscillator.

The second excited state, labeled 2, has non-zero response 
to the $x^2$-perturbation only and, as seen from Fig.~\ref{fig:dens0},
has as expected a two-node gerade (even) profile.
It originates from a single-particle excitation 
of the relative motion and, therefore, 
strongly depends on the inter-particle interaction strength. 
This state is also well described 
by the LR-GP theory.

At this point the predictions of the LR-GP and 
LR-MCTDHB theories start to deviate from each other.
In addition to the widely known excitations of LR-GP, 
we find in our new theory excitations 
with finite response weights at different energies, 
see Fig.~\ref{fig:h1_N100_N10_ho}.
The LR-GP predicts the third excitation at energy 
of about $3\omega_{ho}$, 
while the LR-MCTDHB(2) theory predicts 
an intense excitation with energy slightly above $2\omega_{ho}$. 
We label this excitation, 
which is not present in LR-GP spectrum,
as 2' because it has zero response to $x$-perturbation, 
and has a two-node, gerade density-response profile 
very similar to the above-discussed excitation labeled 2.
In Fig.~\ref{fig:dens0} for better visibility, 
the density-response profile corresponding to excitation 2' is 
magnified five times.

To get a deeper insight into the nature of this excited state, 
we analyze the type of LR-MCTDHB excitation involved.
Since any eigenvector of the response matrix is normalized, 
by using Eq.~\eqref{eq:type} one can compute 
the relative contribution of the orbitals' and CI components, $t_k$.
We recall for reference that all LR-GP excitations
are orbital-like, i.e., $t_k=0$.
In the upper panel of Fig.~\ref{fig:h1_N100_N10_ho} 
we plot by the gray bars
the results of such a decomposition analysis 
for each LR-MCTDHB excitation.
We will discuss the general usefulness of these 
quantities in details in Sec.~\ref{subsec:h10}, 
here we use them first to analyze the nature of the 2' excitation.
In Fig.~\ref{fig:h1_N100_N10_ho} one can see that 
the 2' excitation is CI dominated.
Moreover, it has a two-hole--two-particle (2h-2p) structure, 
i.e., two atoms are excited from the initially condensed GP-like state, 
making thereby two holes (2h) there, 
to the lowest excited orbital, 
thus creating two particles (2p) in this orbital.
This state, therefore, represents a higher harmonic 
of the dipole excitation in a harmonic potential.
Here we show that with a many-body ansatz for the ground state,
our LR-MCTDHB theory provides 
direct access to this excitation.
We discuss the convergence of this excitation below.

In general, all 2h-2p excitations have smaller response 
weights and smaller response of the position-space densities,
compared to their one-hole--one-particle (1h-1p) counterparts. 
To prove this statement let us consider a trapped non-interacting system, 
its ground $|N,0\ra$, 
and its lowest-in-energy 1h-1p and 2h-2p excited states $|N-1,1 \ra$ and $|N-2,2 \ra$. 
The response of the system to a perturbation $f(x)$ is defined
by a superposition of individual contributions from 
every excited state available in the system with weight $\gamma$.
For a non-interacting system the weight of the 1h-1p 
excitation $|N-1,1 \ra$ can be computed as:
$\gamma=\la N,0|\hat a^\dagger _1 \hat a_2 |N-1,1 \ra \int \psi_1^*(x) f(x) 
\psi_0(x) dx=\sqrt{N} \int \psi^*_1(x) f(x) \psi_0(x) dx$.
Here, $\psi_i(x)$ are the eigenfunctions of the trap potential.
We see that 
the intensity of any 1h-1p excitation 
is proportional 
to the square root of the number of particles $\sqrt{N}$ times a transition integral.
This result should be contrasted with that of a 2h-2p excitation 
which contributes with zero intensity because 
$\la N,0|\hat a^\dagger_1 \hat a_2|N-2,2 \ra\equiv0$.
We can thus conclude that 
the non-zero linear-response 
weight of the 2' 
excitation is solely due to 
inter-particle interactions.

Let us discuss the last, higher-energy 
part of the spectra presented in Fig.~\ref{fig:h1_N100_N10_ho}.
Our LR-MCTDHB(2) theory predicts two lines, 
marked as 3' and 3 respectively, 
while the LR-GP only one line 
labeled as 3.
This attribution is based on one and the same analysis scheme as done above,
namely,
that the excitations are numerated according 
to the underlying nodal structures (number of nodes) 
in the respective response density. 
From Fig.~\ref{fig:dens0} it is clearly seen that the density response 
of these lines has a three-node, ungerade shape. 
As a consequence, these states do not respond to the $x^2$-perturbation.
The excited state marked as 3 is presented 
in both the LR-GP and LR-MCTDHB computations
and can be visualized 
as a 1h-1p excitation
of one boson 
from the ground to a third excited orbital.
The additional excitation, labeled as 3',
found by LR-MCTDHB is a non-mean-field-based state, 
because it corresponds 
to a two-boson excitation from the ground state. 
It has an interesting structure --
one boson is excited to the lowest one-node ungerade orbital, 
and the second boson is simultaneously transferred 
to the lowest-in-energy two-node gerade orbital.
We discuss the convergence of this excitation below.

Finally, let us compare the LR-MCTDHB(2) and LR-GP 
spectra for the system made of $N=100$ particles. 
We have chosen
the inter-particle interaction such that the 
non-linear parameter $\lambda_0 N=1$ is the 
same as in the above-discussed case of $N=10$ bosons.
This choice guarantees that the GP and LR-GP 
solutions for both systems are essentially the same.
By comparing the 
LR-GP spectra for $N=10$ and $N=100$ 
depicted in Fig.~\ref{fig:h1_N100_N10_ho}
one can see that, indeed, the positions of the 
spectral lines and their relative intensities 
are the same in 
both cases.
The only difference is the absolute 
value of the intensities, 
which, as it follows from the above discussion,
are proportional to the square root of the 
number of particles $\sqrt{N}$ times a transition integral.
The LR-MCTDHB results for the mean-field-based (1h-1p) excitations, 
labeled as 1,2,3 are quite similar to the LR-GP ones
-- their absolute intensities follow the same $\sqrt{N}$ scaling law.
The main difference
is found in 
the intensities of the 2h-2p excitations. 
Indeed, one can clearly see from Fig.~\ref{fig:h1_N100_N10_ho} 
that the 
relative intensity 
of the 2' line in the $N=100$ 
spectrum is smaller than that in the $N=10$ case.
This means that with the same 
one-body perturbing field
and for the same interaction parameter $\lambda_0 N$,
it is more difficult to excite the 2'
excitation in a system with 
a larger number of particles.

Summarizing, we have contrasted and analyzed the predictions 
of the standard LR-GP (BdG)
and our many-body LR-MCTDHB(2) theories in the situation
where the initial state is essentially completely condensed, i.e.,
the GP and LR-GP theories are believed to provide 
adequate descriptions.
The many-body LR-MCTDHB theory contains the mean-field-based excitations, 
and also predicts additional many-body excited states
which are out of 
the realm of the mean-field linear response.
The response of these many-body excited states 
to the perturbations studied strongly depends on the details of 
the system -- 
on the inter-particle interactions and on the total number of atoms.

\subsubsection{Including more modes (N=10 and M=4,5)\label{subsec:h10}}

\begin{figure}[ht]
\includegraphics[width=.65\columnwidth,angle=-90]
{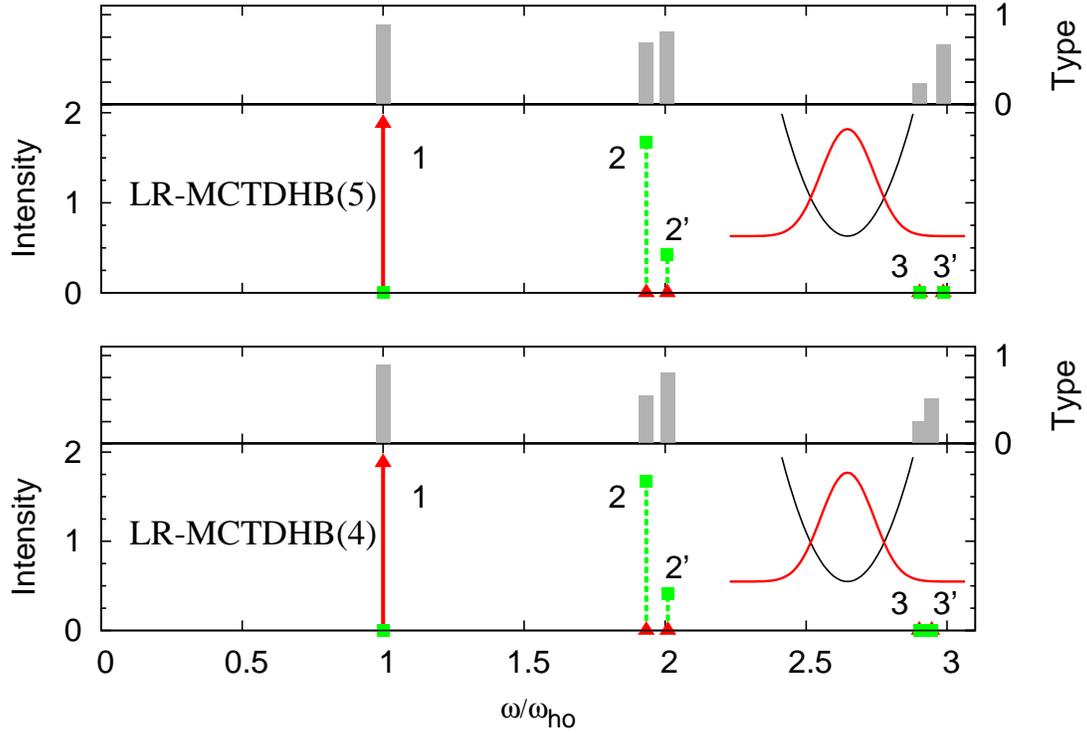} 
\caption{(Color online) Convergence of the LR-MCTDHB(M) theory. 
Comparison 
of the linear-response spectra 
for $N=10$ bosons in a one-dimensional harmonic potential subject 
to linear 
and quadratic perturbations computed 
at the LR-MCTDHB(4) and LR-MCTDHB(5) levels of theory.
The depicted response weights and state 
characterizations are done in the same way 
as in Fig.~\ref{fig:h1_N100_N10_ho}.
The main observation is that the lowest-in-energy 
excitations are numerically converged already 
at a low level of the LR-MCTDHB($M$) theory.
By increasing $M$ we improve the quality of 
description and converge the higher-energy excitations as well,
see text, table \ref{TT}, and Fig.~\ref{fig:R} for further details.
All quantities are dimensionless.
\label{fig:h1_N10_MCTDHB4_vs_MCTDHB5}}
\end{figure}

Now, having investigated many-body excitations in 
a harmonic potential,
we would like to address the question how reliable the obtained predictions are.
To answer this question we wish to study the convergence of the results.
Namely, we study now how the inclusion of more 
orbitals in the LR-MCTDHB theory will change the linear-response results.
Since computation of the linear response with 
more than two orbitals is a more involved numerical task,
we restrict ourselves here to studying of  
a smaller system with $N=10$ bosons only. 
In Fig.~\ref{fig:h1_N10_MCTDHB4_vs_MCTDHB5} 
we present the linear-response spectra 
computed for this system 
at the LR-MCTDHB(4) and LR-MCTDHB(5) levels of theory.

It is worthwhile to mention that the MCTDHB($M$) 
theory is capable of providing numerically-exact description of the statics
and dynamics of interacting many-boson systems 
\cite{Benchmarks,3well,sakmann:09}.
So, by increasing the number $M$ of the 
self-consistent (time-adaptive) modes used in the computations
we increase the quality of description of the ground-state wave-function. 
Since the ground-state wave-function 
is used by the linear-response theory to access the excited states,
the quality of the excited states also improves with 
increasing $M$.
Indeed, by comparing the LR-MCTDHB($M$), $M=1,2,4,5$ 
excitation spectra depicted in Fig.~\ref{fig:h1_N100_N10_ho} 
and Fig.~\ref{fig:h1_N10_MCTDHB4_vs_MCTDHB5} 
we observe the convergence of the results.
Let us discuss the convergence of the excitations.
To aid our discussion and to clearly visualize the
convergence of excitations we collect
from Fig.~\ref{fig:h1_N100_N10_ho} and Fig.~\ref{fig:h1_N10_MCTDHB4_vs_MCTDHB5} 
in table \ref{TT} the frequencies of the 1,2,2',3,3' excitations as
a function of $M$.
The intensities of the lines follow a similar trend concerning the
convergence.

The first two excitations,
1 and 2, 
are available already at the BdG level [LR-MCTDHB($M=1$)].
Table \ref{TT} shows that excitation 1 is already converged at the BdG level
and that excitation 2 converges nicely with $M$.
The third excitation, 2', is first uncovered at the LR-MCTDHB($M=2$)
level, i.e., it cannot be found within BdG theory.
Table \ref{TT} shows that it converged nicely with $M$.
The physical nature of this excitation
is further discussed below.
The next two excitations,
3 and 3', change their relative position between Fig.~1 and Fig.~3,
which makes it harder to see that they converge.
This is particularly the case for excitation 3',
which is first uncovered at the LR-MCTDHB($M=2$) level.
By examining table \ref{TT} we see the convergence of these excitations as well.

\begin{table}
\begin{tabular}{|c|c|c|c|c|c|c|}
\hline
  Excitation \#       &    M=1             &     M=2              &    M=4               &     M=5              &  Type  &  Exact analytical  \\ \hline\hline
      1           &   1.00000000   &     1.00000000   &    1.00000000     &     1.00000000   &   CM   &   1.00000000   \\ \hline
      2           &   1.93782050   &     1.93549034   &    1.93344998     &     1.93335009   &   rel    &         \\ \hline
      2'           &    n/a              &     2.04948358   &    2.00990796     &     2.00785285   &   CM   &  2.00000000    \\ \hline
      3           &   2.90297701   &     2.90431103   &    2.90163006     &     2.90254983   &   rel    &        \\ \hline
      3'           &    n/a              &     2.81478928   &    2.94532236     &     2.98629877   &   CM   &  3.00000000     \\ \hline
\end{tabular}
\caption{Frequencies of the excitations of $N=10$ bosons in the harmonic potential 
(in units of the trap frequency $\omega_0$) as a function of $M$. 
The results are taken from Figs.~1 and 3 of the manuscript.
Convergence of LR-MCTDHB is now clearly seen.
Excitations 2' and 3' are first uncovered at the LR-MCTDHB($M=2$)
level, i.e., they cannot be found (n/a) within BdG theory.
The type of excitations -- `CM' and `rel' -- is indicated.
The exact frequencies of the `CM' excitations are analytically known.}
\label{TT}
\end{table}


From table \ref{TT} we see that the higher excited 3 and more so 3' states converge slower as
a function of $M$ than the lower excitations.
This is a general tendency expected for any many-body method.
Yet, there is more to that than merely energetics.
In the harmonic trap,
the physical nature of the excitations is either `CM' or `rel' excitations.
Excitation 3' (as is excitation 2') is a `CM' excitation, 
a higher harmonic of excitation 1.

\begin{figure}[ht]
\includegraphics[width=.65\columnwidth,angle=-90]
{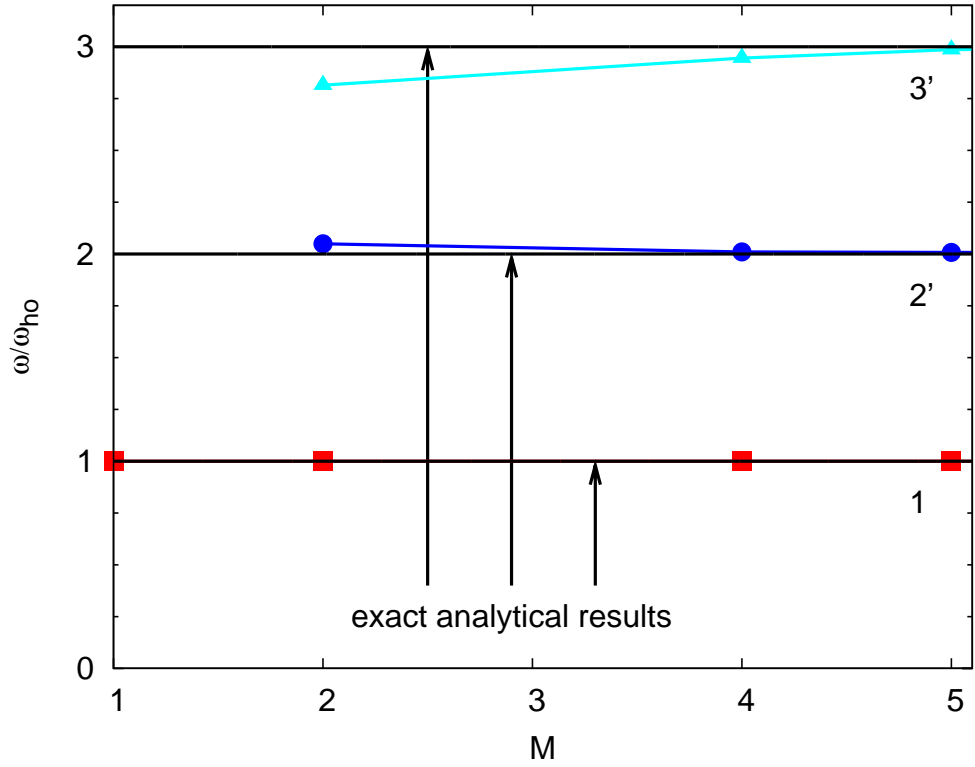}
\caption{Convergence of the center-of-mass excitations of $N=10$ bosons in the harmonic potential as a function of $M$
(data combined from Figs.~\ref{fig:h1_N100_N10_ho} and \ref{fig:h1_N10_MCTDHB4_vs_MCTDHB5}).
The horizonthal lines are the exact analytical values.
All quantities are dimensionless.
\label{fig:R}}
\end{figure}

Generally, the `CM' excitations converge slower than the `rel' excitations, see table \ref{TT}.
There are two reasons for that:
(i) except for the fundamental excitation 1,
these excitations are not described at the $M=1$ (BdG)
level, so to start with one needs more orbitals to describe them, and
(ii) they consist of excitations of a single collective particle.
The MCTDHB ground-state wave-function is constructed in the laboratory frame,
and requires quite a few self-consistent orbitals to faithfully 
represent the `CM' coordinate
for these higher excitations.
Fortunately, the energies of the  
`CM' excitations are analytically known -- they are (in units
of the trap frequency $\omega_0$) just the integers $1,2,3,...$,
which allows us to assess their convergence
(the `rel' excitations depend on the interaction strength 
and are generally not analytically known).
Fig.~\ref{fig:R} depicts as a function of $M$
the frequencies of the `CM' excitations and
follows their convergence -- 
to the analytically known results.

Here it is important to mention that the MCTDHB($M$) method is a variational theory 
\cite{MCHB,streltsov:07,alon:08} -- 
the variational principle guarantees that the approximate 
solution converges to the exact one 
with $M$ (also see \cite{Benchmarks}). 
Moreover, 
the total energy approaches the exact value from the above.
The linear-response part of the LR-MCTDHB($M$) method is not variational, 
implying that the positions of the lines in the linear-response 
excitation spectra can be above as well as below the exact values,
see for an 
example excitations 2' and 3' in Fig.~\ref{fig:R}, respectively.
However, as a general prescription, to get numerically 
converged results for highly excited states within a linear-response 
method one has to provide a very good 
ground many-body wave-function to start from.

Let us now address a general aspect of the LR-MCTDHB theory, namely, 
the above-used decomposition of the LR-MCTDHB excitation 
eigenvectors into orbitals' and CI components, 
according to Eq.~\eqref{eq:type}. 
Clearly, if the number $M$ of self-consistent
orbitals used tends to infinity, 
then one gets the 
exact results already 
within the CI part, 
because the complete Fock space is spanned there.
In this case there is no need to use self-consistent (time-adaptive) basis sets. 
Within the above-defined nomenclature for orbitals' and CI contributions, 
it would mean that the orbital part of the 
linear-response equations 
does not contribute at all. 
Conversely, a non-zero contribution from the orbital 
part means that 
self-consistency is still desirable.
Indeed, a close inspection of the the gray bars 
plotted in Fig.~\ref{fig:h1_N100_N10_ho} 
and Fig.~\ref{fig:h1_N10_MCTDHB4_vs_MCTDHB5} supports this conjecture.
In this respect it is especially interesting and instructive 
to analyze the 3 and 3' excitations 
of the LR-MCTDHB(2) spectrum.
They have structures, involving excitations to the 
one-particle orbitals with two and three nodes, 
see Fig.~\ref{fig:dens0}.
However, within the Fock subspace spanned by 
the MCTDHB($M=2$) theory the needed CI excitations 
are in principle unavailable, 
because all permanents are constructed by permuting 
bosons among two single-particle functions (orbitals).
Neither of these orbitals has the 
required two- or three-node structure: 
the first orbital has no nodes, 
the second orbital has only a one-node profile.
Nevertheless, the linear-response on-top of the MCTDHB(2) 
theory provides us with these excitations.
As expected, they have the orbital-dominating 
structure, i.e., the contribution 
from the CI part is very small.

Concluding, the LR-MCTDHB($M$) equations provide 
improved description of the excitations
which are poorly spanned by or even completely 
out of the MCTDHB($M$) Fock subspace.
The LR-MCTDHB($M$) provides converged results on excited states with 
increasing $M$. 
To get better excited states one has to 
provide a better initial state. 

\subsection{Applications to Bose-Einstein condensates 
in shallow double-well potentials\label{sec:dw}}

In the preceding discussions we have seen that 
even in a simple harmonic well, 
where the BEC exhibits small fragmentation,
our many-body response theory predicts new excitations, 
not described by the standard LR-GP approach.
We now calculate the many-body excitation spectra 
of Bose-Einstein condensates trapped in
a shallow symmetric one-dimensional double-well potential
\be\label{eq:pot_d}
V(x)=b/2\cdot\cos{(\frac{\pi}{3}x)}+\omega_{ho}^2 x^2/2\,,
\ee
with $b=5$ and $\omega_{ho}=\sqrt{2}$, 
for the 
same systems of $N=10$ and $N=100$ bosons with $\lambda_0 N=1$.
This is an intricate, delicate and interesting problem, 
because in this regime of the parameters 
the double well is so shallow that the spatial modes 
are still spatially strongly overlapping,
excluding thereby 
the applicability of, e.g., 
the Bose-Hubbard theory.
Remarkably, the ground-state fragmentation 
in this case is still very small -- it is
about $0.2\%$ and  $0.03\%$ for the systems 
of $N=10$ and $N=100$  bosons, respectively.
Hence, the popular LR-GP theory is the main source 
of information on excited states available in such systems.
Let us see how the LR-MCTDHB results change 
the BdG excitation picture.

\subsubsection{Results for N=100 and N=10\label{subsec:d100}}

\begin{figure}[ht]
\begin{tabular}{ll}
\vspace{-1.5cm}
\includegraphics[width=.55\columnwidth,angle=-90]
{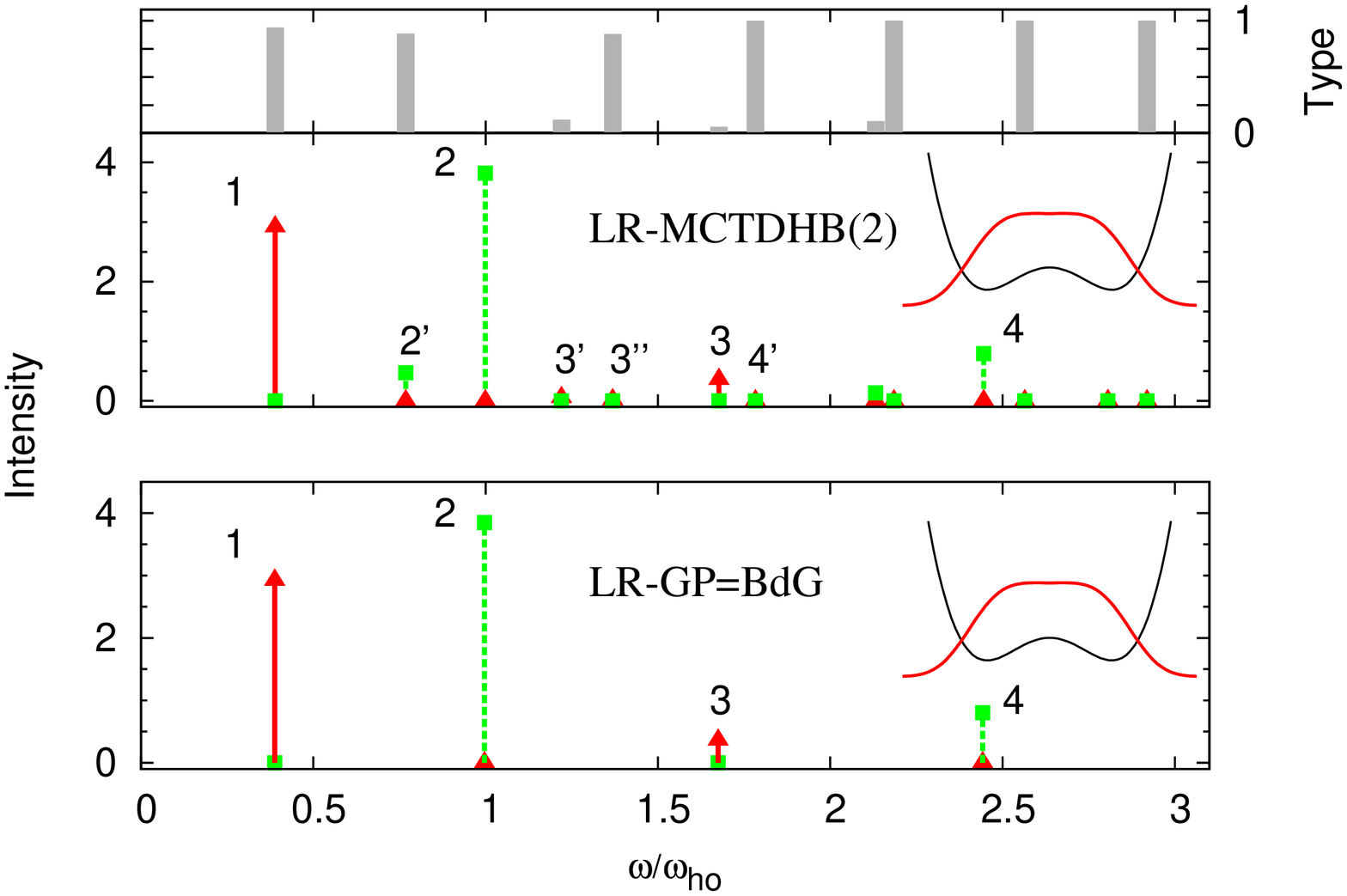}\\           
\vspace{-1cm}
\includegraphics[width=.55\columnwidth,angle=-90]
{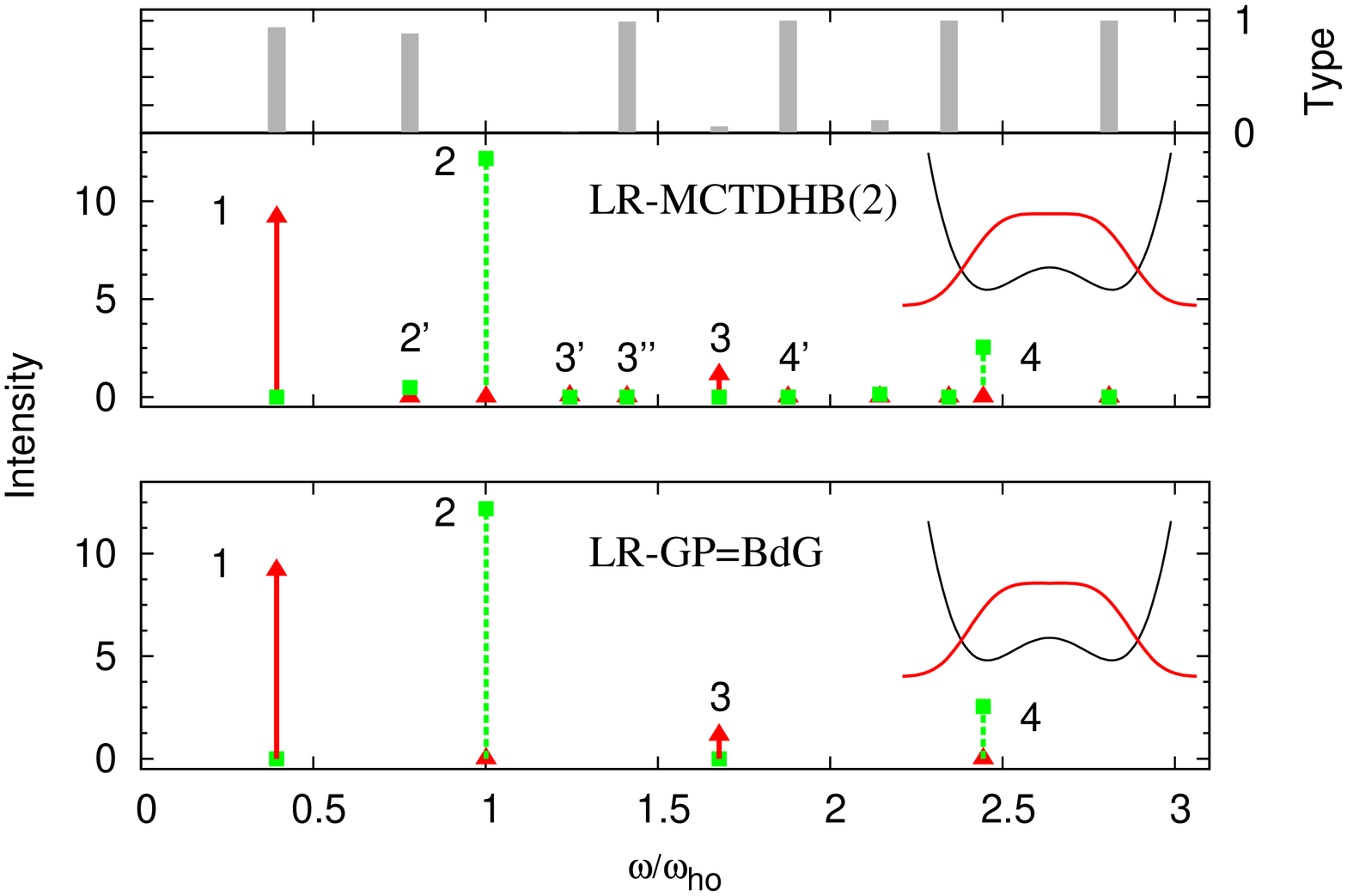} 
\end{tabular}
\vglue -0.5 truecm
\vspace{0.75cm}
\caption{(Color online) Comparison 
of the standard LR-GP (BdG) and LR-MCTDHB(2) linear-response spectra computed 
for the systems of $N=10$ (upper two panels) and $N=100$ (lower two panels) 
bosons
trapped in a very shallow one-dimensional double-well potential, Eq.~\eqref{eq:pot_d},
and subject to 
linear and quadratic perturbations. 
The trap potential and ground-state density (red solid line) 
are shown in the insets.
The depicted response weights
and the scheme of the state enumeration and characterizations 
are the same as in the harmonic case, 
see Fig.~\ref{fig:h1_N100_N10_ho} and text.
The key observation here is that the number of 
low-lying excitations appearing on the many-body level and 
not described by the BdG theory
is much larger in comparison with 
the harmonic-trap results.
All quantities are dimensionless.
\label{fig:h1_N100_N10_dw}}
\end{figure}

The LR-GP (BdG) and LR-MCTDHB(2) excitation spectra 
for the systems of $N=100$ (lower two panels) and $N=10$ (upper two panels) 
bosons trapped in the shallow double well 
are depicted in Fig.~\ref{fig:h1_N100_N10_dw}.
The GP and MCTDHB(2) densities of the ground state 
are very similar and schematically shown in the 
insets together with the trapping potential.
The ground-state density is, because of the central barrier, 
broader than in the harmonic case, 
but 
there is no spatial separation of the modes 
at this barrier height. 
The x-axes in Fig.~\ref{fig:h1_N100_N10_dw} indicate 
excitation energies in units of the envelop trap's frequency $\omega_{ho}$,
see Eq.~\eqref{eq:pot_d}.
The depicted response weights $\gamma_k$ 
and state characterizations and attributions $t_k$
are computed and performed, respectively,
in the same way as in the above-studied harmonic case.
The symmetry of the double-well trap potential allows us to study 
even- and odd-parity excitations separately,
so we use solid (red) lines with triangles to mark the 
odd excitations and dashed (green) 
lines with squares to depict the even ones. 

\begin{figure}[ht]
\includegraphics[width=.95\columnwidth]
{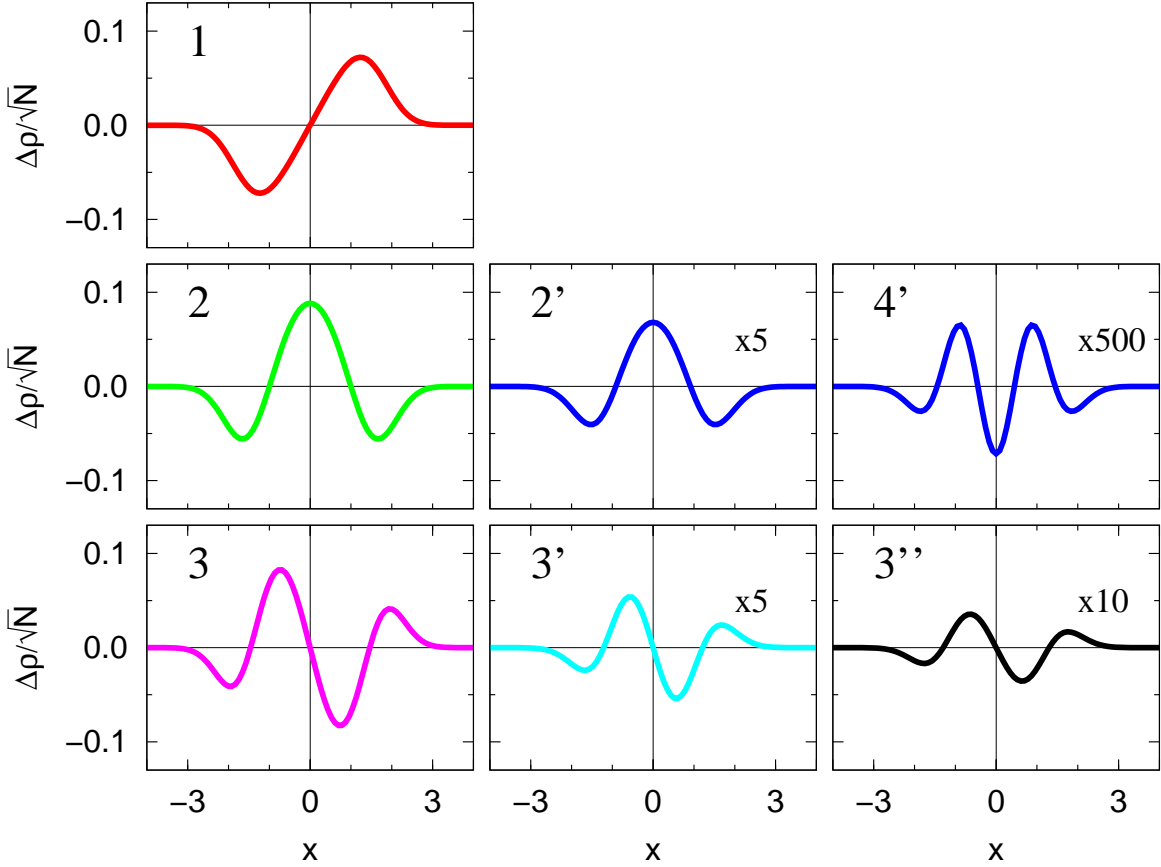}
\caption{(Color online) Density 
responses for $N=10$ atoms in a very shallow 
one-dimensional double-well potential, 
Eq.~\eqref{eq:pot_d}. 
Plotted are the real parts of the response 
densities $\Delta\rho(x)=\Re[\Delta\rho^k_o(x)+\Delta\rho^k_c(x)]$ 
corresponding to real-valued $\gamma_k$'s, 
see discussion in the text. 
All the excitations are numerated according to 
the underlying nodal structures (number of nodes) 
in the respective response density. 
The LR-GP (BdG) results for the first few excitations are depicted 
in the left column 
of the figure. 
The LR-MCTDHB results for these excitations  
are quite similar and, therefore, not shown. 
The many-body excitations,
not describable by the BdG theory,
are labeled 
by the numbers with a prime, 
and magnified for better visibility.
All quantities are dimensionless. 
\label{fig:dens5}}
\end{figure}

We first compare the 
LR-GP and LR-MCTDHB(2) excitation spectra 
for the system of $N=10$ bosons.
The main observation is that the many-body spectrum has 
more spectral features in comparison with 
the LR-GP mean-field-based spectrum.
The second observation is that the spectral lines, 
marked as 1,2,3,4, 
are similarly 
described by both the standard LR-GP and LR-MCTDHB methods.
In Fig.~\ref{fig:dens5} we plot the LR-GP and 
LR-MCTDHB density responses corresponding to 
several low-lying spectral lines including these ones. 
Similarly to the harmonic case, 
the most intense spectral lines can be attributed 
to single-particle (1h-1p) excitations from the condensate to higher excited modes.
The LR-GP and LR-MCTDHB results for these excitations 
have very similar energetics, response weights,
as well as density responses. 

However, there are several very important and significant 
differences between the mean-field-based
and many-body linear-response predictions. 
First of all, 
according to the LR-GP theory the lowest-in-energy excitation of the gerade symmetry 
is at about 1 unit of energy and corresponds 
to a single-boson transfer from the condensate to the second-excited two-node mode.
In contrast, our many-body theory predicts that 
the lowest-in-energy response to the gerade perturbation
takes place at 0.75 units of energy and corresponds 
to a many-body excited state, 
where two bosons are transferred from
the condensate to the one-node mode. 
The next observable difference is that 
in between the mean-field-based excited states labeled as 2 and 3 
there are two many-body excited states labeled as 
3' and 3''.
There are even more excitations lying in the energy 
window between the 
3rd and 4th mean-field-based 
excited states.

The tendency concerning the density responses of two-particle 
excitations observed in the harmonic case
also persists in shallow double-wells. 
In Fig.~\ref{fig:dens5} 
one can see that the non-mean-field-based 
excitations, 
marked as 2' and 3' and corresponding to 
two-boson (2h-2p) excitations, provide smaller
density responses to external perturbations 
in comparison with their 1h-1p counterparts. 
The higher excitations
can also 
be characterized according to their nodal structure, 
see the excitations labeled 3'' and 4' in Fig.~\ref{fig:dens5}.
However, their density responses are 
even weaker -- we have to magnify them significantly 
for better visibility.
We conjecture that other than
one-body response operators will have
to be considered in order to
activate such excitations more efficiently.

Let us now compare the $N=10$ and $N=100$ 
spectra depicted in Fig.~\ref{fig:h1_N100_N10_dw}.
Since we used the same non-linearity $\lambda_0 N=1$ for both systems, 
the position of the LR-GP lines and their relative intensities are the same.
The LR-MCTDHB(2) spectra of these systems, however, 
reveal some differences. 
First of all, 
there are small differences in 
the relative intensity and position of the lowest-in-energy many-body 
spectral line labeled as 2'. 
There are also small differences in the positions 
of the spectral lines corresponding to higher non-mean-field-based excited states.
The differences between the
excitation spectra of the $N=10$ and $N=100$
systems become more pronounced at higher energies.

Summarizing, the LR-MCTDHB theory predicts that 
the excitation spectrum of a condensate in a shallow double well
possesses some additional spectral features not described by the standard LR-GP theory.
In particular, the energy of the lowest-in-energy 
excitation of even symmetry is almost $25\%$ lower than predicted by LR-GP theory.
Moreover, this excitation is not of a single-particle 
nature as predicted by LR-GP theory, 
but it rather consists
of a transfer of two bosons from the condensate to the lowest ungerade mode. 
Generally, in shallow double-well systems
the number of low-lying non-mean-field-based states 
is larger in comparison with the harmonic case. 
The existence of low-lying excited states 
not described by the LR-GP theory can have very important 
consequences on the quantum dynamics and temperature 
properties of ultra-cold systems 
trapped in unharmonic potentials.

\subsubsection{Including more modes (N=10 and M=4,5)\label{subsec:d10}}

\begin{figure}[ht]
\includegraphics[width=.65\columnwidth,angle=-90]
{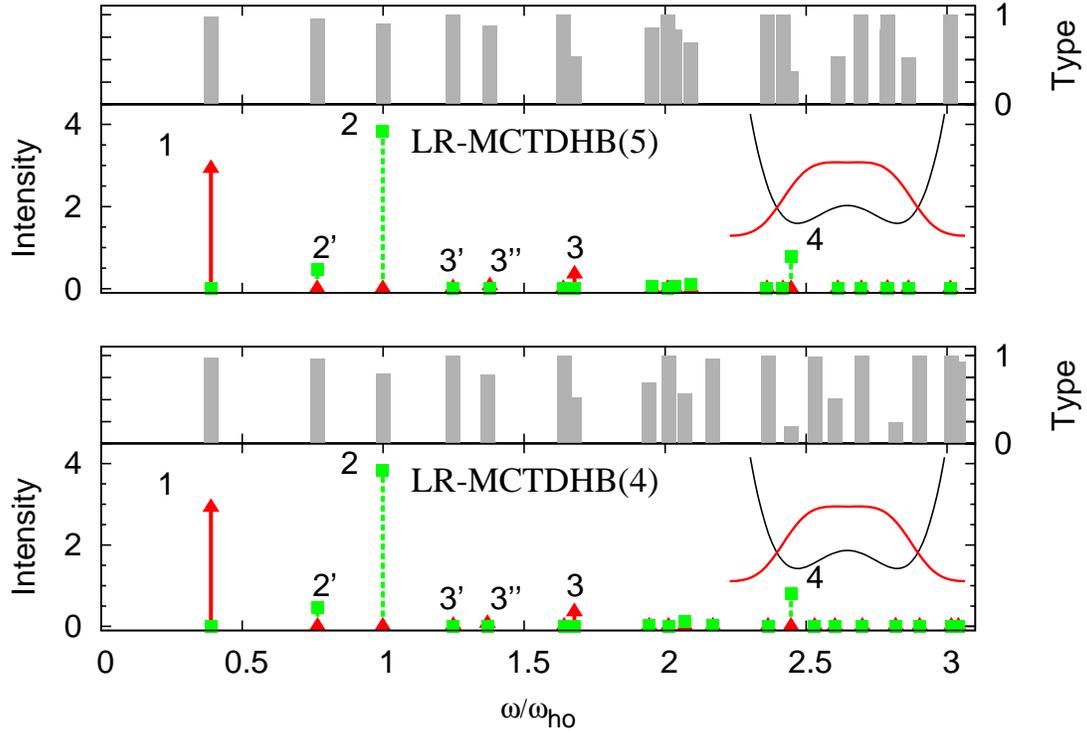}
\caption{(Color online) Convergence of LR-MCTDHB($M$) theory. 
Comparison of the linear-response spectra 
for $N=10$ bosons in a very shallow one-dimensional double-well potential, 
Eq.~\eqref{eq:pot_d},
and subject to linear and quadratic perturbations.
The depicted LR-MCTDHB(4) and LR-MCTDHB(5) response weights 
and state characterizations are 
obtained in the same way 
as in Fig.~\ref{fig:h1_N100_N10_ho}, see text for more details. 
The main observation is that the lowest-in-energy 
excitations are numerically converged already at a 
low level of the LR-MCTDHB($M$) theory. 
By increasing $M$ we improve the quality of description 
of the higher-energy excitations as well.
All quantities are dimensionless.
\label{fig:dw_N10_MCTDHB4_vs_MCTDHB5}}
\end{figure}

Now our goal is to investigate the convergence 
of the LR-MCTDHB($M$) results for double-well traps.
Again, to make the computations more feasible we consider 
the system of $N=10$ bosons and compare the LR-MCTDHB(M) spectra for $M=1,2,4,5$. 
The LR-MCTDHB(1)$\equiv$LR-GP and LR-MCTDHB(2) results 
are depicted in Fig.~\ref{fig:h1_N100_N10_dw}, the LR-MCTDHB(4) and
LR-MCTDHB(5) spectra 
are plotted in Fig.~\ref{fig:dw_N10_MCTDHB4_vs_MCTDHB5}.

In Fig.~\ref{fig:dw_N10_MCTDHB4_vs_MCTDHB5} the low-lying 
parts of the LR-MCTDHB(4) and LR-MCTDHB(5)
spectra are identical,
indicating the convergence of the results. 
As expected, differences start to appear at higher excitation energies and become 
more pronounced in the energy window between 2 
and 3 units of energy.
It is clear that new excitations which
appear due to the inclusion of more
orbitals would need even more orbitals to converge.

Summarizing, the LR-MCTDHB method is capable of providing 
converged results for excited states of a BEC in shallow double-wells.
The LR-MCTDHB method provides better description 
of the low-lying excited states than the highly-excited ones
at a given level of theory $M$;
In order to obtain 
numerically converged results 
for higher excited states one has to perform the computation with a better
initial state, i.e., to use higher levels of 
the MCTDHB($M$) theory.

\section{Summary and concluding remarks\label{sec:con}}

To explore the excitation spectrum of trapped 
interacting bosons
we derived a general 
many-body linear-response theory obtained 
from linearization of the MCTDHB equations.
We have applied the developed LR-MCTDHB theory to study excitations of  
BECs trapped in the harmonic and shallow double-well traps. 

The LR-MCTDHB theory consists 
of a linear-response matrix 
which accounts for the coupled response of the orbitals and state vector. 
This is reflected in the existence of response amplitudes for 
both the orbitals and CI coefficients, 
which are associated with the same excitation frequency. 
Similar to the linear-response 
theories of a fully-condensed (LR-GP, \cite{castin:98})
and fully-fragmented (LR-BMF, \cite{grond:12}) BECs, 
the orbitals' response amplitudes are orthogonal 
to all the ground-state orbitals. 
The response matrix can be divided into submatrices. 
One accounts for the orbitals, and can be considered 
as the generalization of the linear-response matrix of LR-BMF. 
Another one accounts for the CI coefficients and contains 
the CI Hamiltonian in it. 
The coupling between the 
orbitals and the state vector 
in the response matrix is through the off-diagonals. 
The response weights and the density response have also been derived,
and we find that they are sums of the 
orbitals' and coefficients' 
parts.

To obtain excitation spectra we first calculate the ground state.
This is done at a certain level $M$ of the MCTDHB($M$) equations.
Thereafter, we explicitly construct the linear-response 
matrix, Eq.~\eqref{eq:eigL}, and diagonalize it numerically,
solving thereby the eigenvalue problem Eq.~\eqref{eq:evL}. 
The obtained eigenvalues give the excitation energies while
the eigenvectors are used to compute the response weights and density responses. 
To shed more light on the nature of the excitations
we have analyzed the underlying structure of the 
respective eigenvectors in terms of the relative contributions
of the orbitals' and coefficients' 
parts.
In the harmonic potential we analyzed the excitations in terms
of center-of-mass and relative-motion degrees of freedom.

The response matrix which provides
the desired excitation energies does
not depend on the special form of the perturbing field.
The choice of the perturbing field can be utilized
to study the nature of the excited states.
Here, we have chosen $x$ and $x^2$
to distinguish between even/odd and dipole/quadruple transitions.
One can easily envision additional perturbing
fields to characterize further the
spectrum of the excited states under inspection.

We have contrasted the predictions of the standard LR-GP (BdG) 
and our many-body LR-MCTDHB theories in the situation
where the initial ground state is essentially completely condensed, i.e.,
the GP and LR-GP theories are believed to provide adequate descriptions.
The LR-MCTDHB reproduces and improves the 
mean-field-based excitations and, also, 
predicts additional many-body excited states
which are out of 
the realm of the mean-field-based linear response. 
In particular, we were able to calculate
the excitation energy of the higher 
harmonic of the dipole excitation in a 
harmonic trap.
In the shallow double-well system the excitation of the 
same nature turns out to be the lowest-in-energy gerade excitation.
Generally, in shallow double-well systems the number of 
low-lying non-mean-field-based excitations is larger 
in comparison to the harmonic case.
Consequently, 
the existence of low-lying excited states
that are not described by the LR-GP (BdG) 
theory can have very important 
consequences for 
the quantum dynamics and temperature 
properties of trapped ultra-cold bosonic systems,
especially in unharmonic potentials.

We have also assessed the convergence of the 
LR-MCTDHB($M$) equations by direct comparison of the response spectra computed 
at different levels $M=1,2,4,5$ of the theory. 
The convergence is well achieved 
for low-lying excitations.
However, in order to obtain 
numerically converged 
results for higher excited states one
faces a growing 
numerical effort because higher 
levels of the MCTDHB($M$) theory have to be used
to provide the very good ground many-body 
wave-function needed to start from.
Hence, future implementations will be based 
on efficient strategies for solving 
the MCTDHB equations \cite{Mapping}.

In conclusion, we derived a theory for excitation spectra 
of trapped interacting Bose systems 
beyond the available approaches. 
Our first results based on the application of this 
new method suggest that it has a vast perspective
to predict excitations also in BECs 
with attractive interactions or dipolar interactions.

\section*{Acknowledgements}

JG appreciates support from the Alexander von Humboldt Foundation. 
We acknowledge financial support by the DFG also 
within the framework of the Enable fund
of the excellence initiative at Heidelberg University, 
the HGS MathComp, 
and a Minerva Short Term research grant. 
Computation time on the bwGRiD and the
CrayXE6 cluster
Hermit at the HLRS in Stuttgart 
are greatly acknowledged.



\end{document}